\documentclass[authoryear,preprint,review,12pt]{elsarticle}

\usepackage{amssymb}
\usepackage{lipsum}

\usepackage{amsmath}
\usepackage{float}

\usepackage{algorithm}
\usepackage{algpseudocode}

\usepackage{wrapfig}

\usepackage{graphicx}

\usepackage{hyperref}

\usepackage{subfig} 

\usepackage{longtable}
\usepackage{geometry} 

\usepackage{listings}
\usepackage{adjustbox}

\usepackage{xcolor}

\journal{Artificial Intelligence in Medicine}

\begin{document}

\begin{frontmatter}



\title{Enhancing Diagnostic in 3D COVID-19 Pneumonia CT-scans through
Explainable Uncertainty Bayesian Quantification}

\author[first]{Juan Manuel Liscano Fierro}
\affiliation[first]{organization={Maestría en Ciencia de datos, Escuela Colombiana de Ingeniería Julio Garavito},
            city={Bogotá},
            country={Colombia}}

\author[first,second,third]{Héctor J. Hortúa}
\affiliation[second]{organization={Grupo Signos, Departamento de Matemáticas, Universidad El Bosque},
            city={Bogotá},
            country={Colombia}}
\affiliation[third]{organization={Data Science Research Centre, Liverpool John Moores University},
            city={3 Byrom Street, Liverpool L3 3AF},
            country={UK}}

\begin{abstract}
Accurately classifying COVID-19 pneumonia in 3D CT scans remains a significant challenge in the field of medical image analysis.  Although deterministic neural networks have shown promising results in this area, they provide only point estimates outputs yielding poor diagnostic in clinical decision-making. In this paper, we explore the use of Bayesian neural networks for classifying  COVID-19 pneumonia in 3D CT scans providing uncertainties in their predictions. We compare deterministic networks  and their Bayesian counterpart, enhancing the decision-making accuracy under uncertainty information.  Remarkably, our findings reveal that lightweight architectures achieve the highest accuracy of 96\% after developing extensive hyperparameter tuning. Furthermore, the Bayesian counterpart of these architectures via  Multiplied Normalizing Flow technique kept a similar performance along with calibrated uncertainty estimates. Finally, we have developed a 3D-visualization approach to explain the neural network outcomes based on SHAP values. We conclude that explainability along with  uncertainty quantification will offer better clinical decisions in medical image analysis, contributing to ongoing efforts for improving the diagnosis and treatment of COVID-19 pneumonia.
\end{abstract}




\end{frontmatter}




\section{Introduction} \label{introduction}

The rapid integration of artificial intelligence (AI) into the medical field, particularly in the domain of clinical imaging, has led to significant advancements in diagnostic capabilities~\cite{shenavarmasouleh2023deeplearninghealthcareindepth}. However, despite these technological strides, one of the major challenges that persists is ensuring the reliability and transparency of AI-driven diagnostic tools, especially in high-stakes environments like healthcare~\cite{arsenos2022large,arsenos2023data}. In the context of COVID-19 pneumonia, accurate classification of 3D CT scans is crucial for effective patient management, but existing methods often lack the ability to adequately quantify the uncertainty in their predictions—a critical factor in clinical decision-making~\cite{li2024advancingcovid19detection3d}. Traditional neural networks, widely employed for medical image analysis, have achieved impressive accuracy in various diagnostic tasks~\cite{kollias2020deep,kollias2021mia,kollias2023deep,hou2022cmc_v2}. Nonetheless, these models typically operate in a deterministic manner, providing single-point estimates without indicating the level of confidence in their predictions. This absence of uncertainty quantification can lead to overconfident and potentially misleading decisions, which is particularly concerning in medical settings where the cost of error is high~\cite{ZAHARI2024108324}. To address this limitation, recent research has explored the use of Bayesian neural networks, which offer a probabilistic framework that not only produces predictions but also quantifies the associated uncertainty. Several studies have proposed methods such as Monte Carlo Dropout \cite{Zhao_2024} \cite{article} \cite{zou2023reviewuncertaintyestimationapplication} and Variational Inference \cite{hu2023interrateruncertaintyquantificationmedical} to approximate Bayesian inference in neural networks, demonstrating their potential to improve the interpretability and reliability of AI models in medical imaging. However, these approaches often involve trade-offs, such as increased computational complexity and potential overestimation of uncertainty, which require further investigation~\cite{guo2017calibration,kristiadi2020bayesianjustbitfixes}. In this research, we aim to compare the effectiveness of deterministic neural networks and Bayesian neural networks in the classification of COVID-19 pneumonia using 3D CT scans. Our approach focuses on leveraging the strengths of Bayesian methods to enhance uncertainty quantification, thereby refining the clinical decision-making process. The main objective of this study is to determine whether Bayesian neural networks can provide more reliable and interpretable predictions compared to traditional neural networks, particularly in the context of medical imaging. The authors in~\cite{mnf} have claimed  that Bayesian neural networks with Multiplicative Normalising Flows(MNF) 3D-layers demonstrate superior performance in terms of uncertainty quantification, providing more actionable insights. These findings highlight the potential of Bayesian approaches to not only improve diagnostic accuracy but also to offer a more nuanced understanding of the confidence in AI-driven diagnoses, ultimately contributing to safer and more effective clinical practices. Finally, we have developed a 3D- explainable routine based on Shapley values~\cite{NIPS2017_7062} to  understand deep learning models decision by highlighting the features deemed relevant by the model. It allowed us to observe regions on the scans relevant for classifying the degree of COVID-19 pneumonia. The paper is developed as follows, in section~\ref{sec2:bayesian_nn}  we explain the
concepts behind of Bayesian Neural Networks and describe briefly some of the main techniques used in this paper for adding uncertainty in the models, in section~\ref{sec3:calibration}  we
have a glimpse about calibration methods and metrics for evaluating the accuracy in the uncertainty estimates and section~\ref{sec4:metrics}  we provide information about the metrics used for evaluating the performance of the model predictions. Section~\ref{sec5:data} describes the dataset used and the methodology is commented in section~\ref{sec6:metho}. Results are shown in section~\ref{sec7:results} and finally the discussion along with conclusions are mentioned in section~\ref{sec8:conc}.

\section{Related work} \label{background}

The need to enhance the reliability and trustworthiness of AI models in healthcare, especially in medical image analysis, is a critical challenge \cite{article_medical_image_classification, Zhao_2024, article_bayesian_conv_nn_medical_imaging}. While AI-driven diagnostic tools have made significant progress, the persistent issue of uncertainty and potential unreliability in model predictions remains a concern for clinical decision-making \cite{zou2023reviewuncertaintyestimationapplication, hu2023interrateruncertaintyquantificationmedical, abboud2024sparsebayesiannetworksefficient}. Bayesian neural networks (BNNs) offer a promising solution by providing probabilistic interpretations of predictions, allowing clinicians to better assess prediction confidence \cite{article_optimization_performance_evaluation}.

The COVID-19 pandemic has brought renewed attention to medical image analysis, particularly in the classification of COVID-19 pneumonia using chest CT scans. Deep learning techniques, such as convolutional neural networks (CNNs), have shown great promise in this area, with successful applications in detecting intracranial hemorrhages \cite{Sharrock2021, Chang1609} and diagnosing COVID-19-related lung conditions \cite{nn_lung_lesions}. However, deterministic neural networks, while effective, do not account for the uncertainty inherent in medical imaging, a limitation that BNNs can overcome.

Recent studies underscore the potential of deep learning in improving diagnostic accuracy and streamlining clinical workflows \cite{LUNDERVOLD2019102, doi:10.1146/annurev-bioeng-071516-044442}. For instance, McKinney et al. \cite{McKinney2020} demonstrated that deep learning algorithms could detect breast cancer with accuracy comparable to experienced radiologists. Similarly, Ardila et al. \cite{Ardila2019} highlighted the effectiveness of AI models in detecting lung cancer from CT scans, showing potential for enhancing global cancer screening.

Among the deep learning techniques, BNNs are particularly valuable in medical imaging due to their ability to quantify uncertainty, a crucial factor in clinical decision-making. This uncertainty quantification enables more informed judgments about the reliability of diagnoses and supports more nuanced treatment decisions \cite{papamarkou2024positionbayesiandeeplearning}.

\section{Bayesian Neural Networks} \label{sec2:bayesian_nn}

Bayesian neural networks (BNNs) are a type of neural network that incorporates Bayesian probability theory into the training and inference process. Unlike deterministic neural networks, which generate a fixed prediction given a set of inputs, BNNs generate a probability distribution over possible predictions~\citep{chai2018uncertainty}.  In~\cite{chai2018uncertainty} discusses the importance of decomposing predictive uncertainty by arguing that random and epistemic uncertainties tell us about different facets of an input value. 
BNNs use prior distributions over the network weights and learn a posteriori distributions $p(w/D)$ using Bayes' rule: $ p(w|D) \sim p(D|w)p(w)$, where $p(D|w)$ denotes the likelihood function representing the probability of the observed data $D$ given the weights $w$, and $p(w)$ representing the prior distribution of the weights. The posterior distribution represents the uncertainty in the weights given the observed data $D = (X, Y)$ is usually approximated by variational inference methods \cite{chai2018uncertainty, heek2018well} or Markov chain Monte Carlo (MCMC) \cite{neal2012bayesian}. Then, the posterior distribution is used to make predictions by taking the expectation of the ouput over the distribution. Once the calculation of the posterior distribution has been performed, the probability distribution of a new test example $x^{*}$ can be determined by~\cite{Gal2016UncertaintyID}

\begin{equation} \label{eq:posterior_test_example}
p(y^{*} | x^{*}, D) = \int_{w}^{} p(y^{*} | x^{*}, w) p(w|D) dw,
\end{equation}	
where $p(y^{*} | x^{*}, w)$ is the posterior predictive distribution corresponding to the set of weights. An advantage of BNNs is that they can provide more robust and reliable predictions by accounting for uncertainty in the data and model~\cite{gal2016dropout}. This is particularly useful in applications where incorrect or uncertain predictions can have serious consequences, such as medical diagnoses or financial forecasts. Another advantage is that BNNs can provide a measure of uncertainty for each prediction, which can be used to guide decision making or to identify areas where more data or model refinement is needed.
\subsection{Variational Inference}
Variational inference approximates the complex posterior distribution $p(w|D)$ with a simpler tractable distribution over the model weights, $q(w)$, with variational parameters $v$. These variational parameters $v$ are fitted so that $q(w)$ approximates the desired posterior distribution $p(w|D)$. This fitted variational distribution is used to make the model predictions instead of the true posterior \citep{chai2018uncertainty}.
	
One way to measure the distance between the two probability distributions $q(x)$ and $p(x)$ is by using the Kullback-Leibler divergence or KL-divergence. Defined as

\begin{equation} \label{eq:5}
KL(q(x)||p(x)) \equiv \mathbb{E}_{q(x)}  \left[log\frac{q(x)}{p(x)}\right]  = \int q(x) log \frac{q(x)}{p(x)} dx.
\end{equation}

To make the variational distribution $q(w)$ close to the posterior distribution $p(w|D)$, it is required to minimize the KL-divergence between these two distributions, obtaining the following~\cite{kendall2017uncertainties}

\begin{equation} \label{eq:9}
    \log p(D) \geq KL(q(w) || p(w)) + \mathbb{E}_{q(w)} [\log p(D|w)].
\end{equation}


Therefore, finding the variational parameters $v$ to minimize KL-divergence is the same as maximizing ELBO. In effect, variational inference translates the problem of inference over the weights distribution into the optimization problem of maximizing the ELBO. Once the objective is defined Once the ELBO objective is defined, we can sample $q(w)$ and use backpropagation, as in DNNs, to find optimal values of the variational parameters that maximize the ELBO.

\subsubsection{Variational Distribution}

The essence of the variational distribution is that it is similar enough to the posterior weight distribution after divergence is minimized, but it is simpler to draw samples. A common form of variational distribution is the mean-field approximation, in which we assume that the variational distribution factorizes into the product of distributions by treating the weights as independent variables. Additionally, we can use a Gaussian distribution as a variational distribution, allowing easier sampling from a normal distribution instead of the exact weight posterior \citep{chai2018uncertainty, mnf, louizos2017multiplicative, Hort_a_2023}. In this case, the variational distribution becomes

\begin{equation} \label{eq:11}
q(w|\theta) = \prod_{ij} \mathcal{N} (w;\mu_{ij}, \sigma^{2}_{ij}),
\end{equation}	

where $i$ and $j$ are the indices of the neurons from the previous and current layer, respectively. Applying the reparametrization trick, we obtained $w_{ij} = \mu_{ij} + \sigma _{ij} * \epsilon _{ij}$, where $\epsilon _{ij}$ was drawn from the normal distribution. Furthermore, if the prior is also a product of independent Gaussians, the KL divergence between the prior and the variational posterior be computed analytically, which makes this approach computationally efficient \citep{gal2016dropout}.

\subsection{Multiplicative normalising flows (MNF)}

The Gaussian mean-field distribution described in the equation Eq.(~\ref{eq:11}) is the most widely used family for posterior variation in BNNs. Unfortunately, this distribution lacks the ability to adequately represent the complex nature of the true posterior. Therefore, it is anticipated that improving the complexity of the variational posterior will produce substantial improvements in performance. This is attributed to the ability to sample from a more reliable distribution, which closely approximates the true posterior distribution. The process of improving the variational posterior requires efficient computational methods while ensuring its numerical feasibility. Multiplicative normalising flows (MNF) have been proposed to efficiently fit posterior distributions by using auxiliary random variables and normalising flows \cite{pmlr-v70-louizos17a}. Mixture normalising flows (MNF) suggest that the variational posterior can be represented mathematically as an infinite mixture of distributions \cite{pmlr-v70-louizos17a}

\begin{equation} \label{eq:12}
q(w|\theta) = \int q(w|z, \theta) q(z|\theta) dz,
\end{equation}	

being $\theta$ the learnable posterior parameter, and $z \sim q(z| \theta) \equiv q(z)$ the vector with the same dimension as the input layer, which plays the role of an auxiliary latent variable. Furthermore, by allowing local reparametrizations, the variation posterior for fully connected layers becomes

\begin{equation} \label{eq:13}
w \sim q(w|z) = \prod_{ij} \mathcal{N} (w;z_{ij}\mu_{ij}, \sigma^{2}_{ij}).
\end{equation}

The flexibility of the variational posterior can be increased by improving the complexity of $q(z)$. This can be done using normalising flows since the dimensionality of $z$ is much lower than the weights. Starting from samples $z{0} \sim q(z{0})$ from fully factorized Gaussians (see Eq.(~\ref{eq:11})), a rich distribution $q(z{K})$ can be obtained by applying successively invertible $f{k}$ transformations,

\begin{equation} \label{eq:14}
z_{k} = NF(z_{0}) = f_{k} \circ \cdots \circ f_{1}(z_{0}),
\end{equation}	

\begin{equation} \label{eq:15}
\log q(z_{K}) = \log q(z_{0}) - \sum_{k = 1}^{K} \log \Big|
    \det \frac{\partial f_{k}}{\partial z_{k-1}} \Big|
\end{equation}

To handle the intractability of the posterior, \cite{pmlr-v70-louizos17a} suggested using Bayes' law again $q(z_{K})q(w|z_{K}) = q(w)q(z_{ K}|w)$ and introduce a new auxiliary distribution $r(z_{K}|w, \phi)$ parameterized by $\phi$, with the purpose of approximating the a posteriori distribution of the original variational parameters $q( z_{K}|w)$ to further lower the bound of the KL divergence term. Consequently, the KL divergence term can be rewritten as follows \citep{pmlr-v70-louizos17a}

\begin{equation} \label{eq:16}
\begin{aligned}[b]
& -KL \left[ q(w) \parallel p(w) \right] \geq \\
& \mathbb{E}_{q(w,z_{k})} \left[ -KL \left[ q(w|z_{k}) \parallel p(w) \right] + \log q(z_{k}) + \log r(z_{k} | w, \phi) \right].
\end{aligned}
\end{equation}

The first term can be calculated analytically since it will be the KL-divergence between two Gaussian distributions, while the second term is calculated by the normalising flow generated by $f_{K}$ (see Eq.(~\ref{eq:15} )). Furthermore, the auxiliary posterior term is parameterized by inverse normalization flows as follows

\begin{equation} \label{eq:17}
z_{0} = NF^{-1}(z_{k}) = g_{1}^{-1} \circ \cdots \circ g_{k}^{-1}(z_{k}),
\end{equation}	

and

\begin{equation} \label{eq:18}
\log r(z_{K}|w, \phi) = \log r(z_{0}|w, \phi) + \sum_{k = 1}^{K} \log \big|
    det \frac{\partial g_{k}^{-1}}{\partial z_{k}}
    \big|,
\end{equation}
where $g_{k}^{-1}$ can be parameterized as another normalising flow. A flexible parameterization of the auxiliary posterior can be given by \citep{pmlr-v70-louizos17a}
\begin{equation} \label{eq:19}
z_{0} \sim r(z_{k}|w, \phi) = \prod_{i} \mathcal{N} (z_{0}; \tilde{\mu_{i}}(w, \phi), \tilde{\sigma_{i}}^2(w, \phi)),
\end{equation}	
where the parameterization of the mean $\tilde{\mu}$ and the variance $\tilde{\sigma}^2$ is carried out by the RealNVP mask as the choice of normalising flows~\cite{dinh2017densityestimationusingreal}.
\subsection{Multiplicative normalising flows in a voxel-grid representation}
\cite{mnf} present the result of the generalization of Eq.(~\ref{eq:13}) towards 3D convolutional layers. Starting with the extension of the variational posterior as
\begin{equation} \label{eq:20}
    w \sim q(w|z) = \prod_{i}^{D_{d}} \prod_{j}^{D_{h}} \prod_{k}^{D_{w}} \prod_{l}^{D_{f}} \mathcal{N}(w; z_{l} \mu_{ijkl}, \sigma_{ijkl}^{2}),
\end{equation}
where $D_{h}$, $D_{w}$ and $D_{d}$ are the 3 spatial dimensions of the boxes, and $D_{f}$ is the number of filters for each kernel. The objective is to address the challenge of improving the adaptability of the approximate posterior distribution of weights coming from a 3D convolutional layer. The algorithm that describes the procedure for forward propagation of each 3D convolutional layer is found in \citep{mnf}.


    

\section{Calibration Methods}\label{sec3:calibration}
\cite{guo2017calibration} states that modern deep neural networks are often uncalibrated. As a result, interpreting predicted numbers as probabilities is incorrect. Real-world problems often require models that produce not only a correct prediction but also a reliable measure of confidence in it. Reliability refers to the estimated probability that the forecast is correct. For example, as \cite{vasilev2023calibration} makes clear, if an algorithm predicts that a given sample of patients is healthy with a confidence of 0.9, we expect that 90\% of them are actually healthy. A model with a reliable confidence estimate is called calibrated. Along with interpreting neural networks predictions, reliable calibration is important when probability estimates are fed into later steps of the algorithm. As noted, many real-life applications require more than just a model that predicts the most likely outcome. This is evident in critical areas such as self-driving cars or medical diagnostics, where poor predictions can lead to significant losses. These applications also require robust and reliable predictions with guarantees on model uncertainty \cite{kendall2017uncertainties}. The importance of using probabilities to express the uncertainty of a prediction has been widely recognized both in classical statistics \cite{dawid1982well,murphy1977reliability} and in Machine Learning environments \cite{Gal2016UncertaintyID, guo2017calibration, hernandez2015probabilistic,li2017dropout}. However, contemporary research lacks a widely recognized benchmark for the quality of predictive uncertainty. Calibration methods offer a powerful tool to evaluate the quality of uncertainty estimation, since well-calibrated predictions can be interpreted as objective probabilities. In the Bayesian framework, a prediction should be considered subjective in the sense that predictions depend on prior assumptions about the behavior of a random process. Objective probability follows the frequentist interpretation of probability, according to which the probability of a random event corresponds to its long-term frequency. \cite{heek2018well} demonstrates how subjective (Bayesian) and objective (frequentist) probabilities can be related using the calibration theorem, which states that any subjective probability model should be considered well calibrated. Additionally, a model with good uncertainty estimates can be used to determine a good balance between exploration and exploitation in boosting algorithms \cite{blundell2015weight}. Current variational inference and sampling methods are not able to outperform simple linear approaches in reinforcement learning based on Thompson~\cite{riquelme2018deep} sampling. 
\subsubsection{Reliability Diagrams} \label{conf_diagrama}
Visual representations of the model calibration can be done via the reliability diagrams. If the model is perfectly calibrated, then the diagram should represent the identity function. Any deviation from a perfect diagonal represents poor calibration. To estimate the expected accuracy of finite samples, the predictions are grouped into $M$ bins of intervals (each of size $1/M$) and the precision of each bin is calculated. Defining $B_{m}$ as the set of indices of samples whose prediction confidence falls in the interval $I_{m} = \left( \frac{m-1}{M}, \frac{m}{M} \right]$. Then, the accuracy of $B_{m}$ is

\begin{equation}\label{eq:acc_bm}
acc(B_{m}) = \frac{1}{\left| B_{m} \right|} \sum_{i \in B_{m}} 1(\hat{y_{i}} = y_{i}),
\end{equation}

where $\hat{y_{i}}$ and $y_{i}$ are the predicted and actual values of the classes for sample $i$. Classical probability says that if $\text{acc}(B_{m})$ is an unbiased and consistent estimator of $\mathbb{P} (\hat{Y} = Y | \hat{P} \in I_{ m})$. The average trust within the container $B_{m}$ would be defined as
\begin{equation}\label{eq:conf_bm}
\text{conf} (B_{m}) = \frac{1}{\left| B_{m} \right|} \sum_{i \in B_{m}} \hat{p_{i}},
\end{equation}
where $\hat{p_{i}}$ is the confidence in sample $i$. Therefore, a perfectly calibrated model will have $\text{acc} (B_{m})$ = $\text{conf} (B_{m})$ for all $m \in {1,...,M }$.
\subsubsection{Expected Calibration Error (ECE)}

While calibration diagrams are very powerful visual tools, it is more convenient to have a statistic that summarizes the calibration evaluation. An indication of poor calibration is the difference in expectations between confidence and accuracy
\begin{equation}\label{eq:diferencias_expectativa}
ECE= \mathbb{E}_{\hat{P}} \left[ \left| \mathbb{P} (\hat{Y} = Y | \hat{P} = p) - p \right| \right].
\end{equation}
ECE approximates Eq.(~\ref{eq:diferencias_expectativa}) by partitioning the predictions into $M$ equally-spaced bins (similar to the definition in section \ref{conf_diagrama}) and takes the weighted average of the difference between accuracy/confidence of containers. More precisely
\begin{equation}\label{eq:ece}
\text{ECE} = \sum_{m = 1}^{M} \frac{\left| B_{m} \right|}{n} \left| \text{acc}(B_{m}) - \text{conf}(B_{m}) \right|, 
\end{equation}	
where $n$ is the number of samples. The difference between acc and conf for a given container represents the calibration \textsl{gap} (usually displayed in the red bars within the reliability diagram).
\section{Metrics to model performance}~\label{sec4:metrics}
When evaluating the effectiveness of classification models, it is essential to employ a diverse range of metrics that provide nuanced information about their performance. These metrics serve as quantitative measures to measure the accuracy, reliability, and generalization ability of the model across different classification tasks. From fundamental metrics like accuracy and precision to more nuanced measures like area under the ROC curve and intercept over union, each metric provides unique insights into the model's strengths and weaknesses. In this section, we delve into a comprehensive examination of several metrics commonly used in evaluating classification models, clarifying their importance, interpretation, and formulas. By comprehensively evaluating models using a combination of these metrics, researchers can gain a comprehensive understanding of their performance and make informed decisions regarding model selection, optimization, and implementation.

\section{Dataset}~\label{sec5:data}
The \cite{mosmed} dataset used in this project consists of anonymous human lung computed tomography (CT) scans with COVID-19-related and not related findings with dimension of $128 \times 128 \times 64$. The scans were obtained between March 1, and April 25, 2020, at medical hospitals in Moscow, Russia. This data as licensed by MosMed serves multiple purposes, including educational material for medical imaging specialists, development and testing of AI-based services, and as a source of information for medical specialists and the general public. The CT scans in the dataset show intrinsic radiological signs of COVID-19 infection. A subset of the studies has been annotated with binary pixel masks, indicating regions of interest such as ground-glass opacifications and consolidations. Although no information is provided on the sex of the patients, each study corresponds to a single patient, and each study is represented by a series of reconstructed images of the mediastinal soft-tissue window. The studies are classified into five groups based on signs of viral pneumonia. The dataset comprises a total of 1110 studies, divided into the following 5 categories:
\begin{enumerate}
    \item CT-0: Represents normal lung tissue without any CT signs of viral pneumonia (254 samples).
    \item CT-1: Indicates several ground-glass opacifications, with less than 25\% involvement of lung parenchyma (684 samples).
    \item CT-2: Depicts ground-glass opacifications with 25-50\% involvement of lung parenchyma (125 samples).
    \item CT-3: Shows ground-glass opacifications with partial consolidation, with 50-75\% involvement of lung parenchyma (45 samples).
    \item CT-4: Illustrates ground-glass opacifications with partial consolidation, with more than 75\% involvement of lung parenchyma (4 samples).
\end{enumerate}
Notice that the distribution of studies into these categories was based solely on radiological findings and not on polymerase chain reaction (PCR) test results or clinical verification. Additionally, given the clinical relevance of distinguishing between the absence and presence of characteristic manifestations of COVID-19 pneumonia, this project has strategically focused on the CT-0, CT-2, and CT-3 classifications for model development and evaluation. By prioritizing these classifications, the project aims to address the major diagnostic challenges faced by clinicians while ensuring clinical relevance and applicability of the findings. CT-1 and CT-4 were considered as an alternative test dataset for evaluating and analysing the results of explainability and uncertainty quantification.

\section{Methodology}~\label{sec6:metho}

Let us start by  extensively investigated the standard methods for processing computed tomography (CT) images, focusing particularly on thoracic images. It was observed that CT images are typically processed using Hounsfield Unit (HU) values in terms of window width and center. Subsequently, we define different HU windows for testing, referenced in Table(~\ref{tab:ventanas_hu}), limiting the pixel values of the images according to the specified HU window width. 
\begin{table}[H]
    \centering
    \tiny  
    \begin{tabular}{|c|c|c|p{0.3\linewidth}|}
        \hline
        ID Window (HU) & Lower Pixel Limit & Upper Pixel Limit & Source \\
        \hline
        W1 & -1000 & 400 & \url{https://keras.io/examples/vision/3D_image_classification/} \\
        \hline
        W2 & -1100 & 500 & \url{https://www.youtube.com/watch?v=totkna0Z-2o\&t=334s} \\
        \hline
        W3 & -950 & 550 & \url{https://www.unsam.edu.ar/escuelas/ciencia/alumnos/PUBLIC.1999-2006-\%20Alumnos\%20P.F.I/(TAC)\%20GUERREIRO\%20MARTINS\%20MARIANO.pdf} \\
        \hline
        W4 & -1000 & 0 & \url{https://www.ncbi.nlm.nih.gov/pmc/articles/PMC7120362/} \\
        \hline
    \end{tabular}
    \caption{HU Windows for Computed Tomography Processing}
    \label{tab:ventanas_hu}
\end{table}
After reading the  Nifti imagefiles (.nii extension), they are followed by a 90 degree rotation to ensure consistent orientation. In order to increase the dataset for a better performance, we use the library \href{https://github.com/ZFTurbo/volumentations}{Volumentations 3D} \cite{solovyev20223d}, where several transformations were implemented for the CT scans, including inverting the X, Y and Z axes, adding Gaussian noise and adjust the color fluctuation to improve the robustness and generalization of the model. The data was normalized in the range 0,1, and it was randomly split  into training (70\%), testing (20\%) and validation (10\%).
We starting to  train several neural network  models  for each HU window. The architecture described in  \href{https://keras.io/examples/vision/3D_image_classification/#make-predictions-on-a-single-ct-scan}{keras-3Dscan} was adopted as a starting point for the classification of 3D images. We then apply data augmentation techniques to these models and compare their performance metrics. We also experiment with diverse popular convolutional neural network architectures available in  \href{https://github.com/ZFTurbo/classification_models_3D}{3D classification}~\cite{solovyev20223d} such as ResNet18-34, SeresNet18-32, EfficientNetB0 and DenseNet121. These models were trained using the best performing HU window identified in the previous step. To explore uncertainty quantification, we build two kind of architectures, one where we replace only the deterministic top layer with Bayesian counterparts, and the second one where we replace all layers  instead (see Fig.~\ref{fig:architecturedet}), leveraging the TensorFlow libraries \cite{tensorflow2015} and \href{https://www.tensorflow.org/probability}{TensorFlow Probability}. We make use of Flipout, reparameterization trick  and multiplicative normalization flows (MNF) techniques  with the purpose of transforming the deterministic model into a probabilistic one with the property of quantifying the uncertainty in their predictions. Finally, as we described earlier, most of the neural networks provide un-calibrated  uncertainties estimates. So, in order to avoid (under)-over estimations, we should apply calibration techniques.  To check how calibrate is the model we use reliable diagrams\footnote{\href{https://github.com/hollance/reliability-diagrams}{Reliable diagrams in python.}} as a tool  to visualize the relationship between predicted probabilities and expected precision. For each \textsl{bin}, the average confidence and accuracy are calculated and plotted against each other. We  report  metrics based on the Expected calibration error (ECE) to quantify the overall performance of the model calibration. It measures the difference in expectations between confidence and accuracy, with lower values indicating better calibration. We also make use of threshold values where the reliability plots and ECE calculations were performed for different cut-off points [0.4,0.8] to   provide information and decision for establishing   a well-calibrated model.
\section{Results}~\label{sec7:results}
Exploring the early stages of the project yielded important insights into model performance and optimization strategies. Here we present a complete description of the results obtained, highlighting the key findings and the methodologies employed. This research embarked on a multifaceted exploration of various methodologies and model architectures to discern optimal approaches for classifying COVID-19 pneumonia in 3D CT scans. This effort encompassed defining the most appropriate Hounsfield Unit (HU) window, evaluating the impact of data augmentation techniques, examining the effectiveness of various neural network architectures, and delving into model calibration and uncertainty estimation with higher performance. Through meticulous analysis, we aim to provide valuable insights into the complexities of model performance and uncertainty quantification in the context of COVID-19 pneumonia classification.
\subsection{HU window} \label{r_ventana_hu}
One of the first findings of the project was the determination of the optimal HU window for CT image processing. Through various experimentations, the HU window, named W4 in Table(~\ref{tab:ventanas_hu}), was found to provide superior performance metrics compared to other windows. In particular, the model trained with window W4 demonstrated the highest accuracy on the testing data set, achieving a score of 0.91 (see Table~\ref{tab:hu_window_selection})

\begin{figure}[H]
    \centering
    \includegraphics[width=1\textwidth]{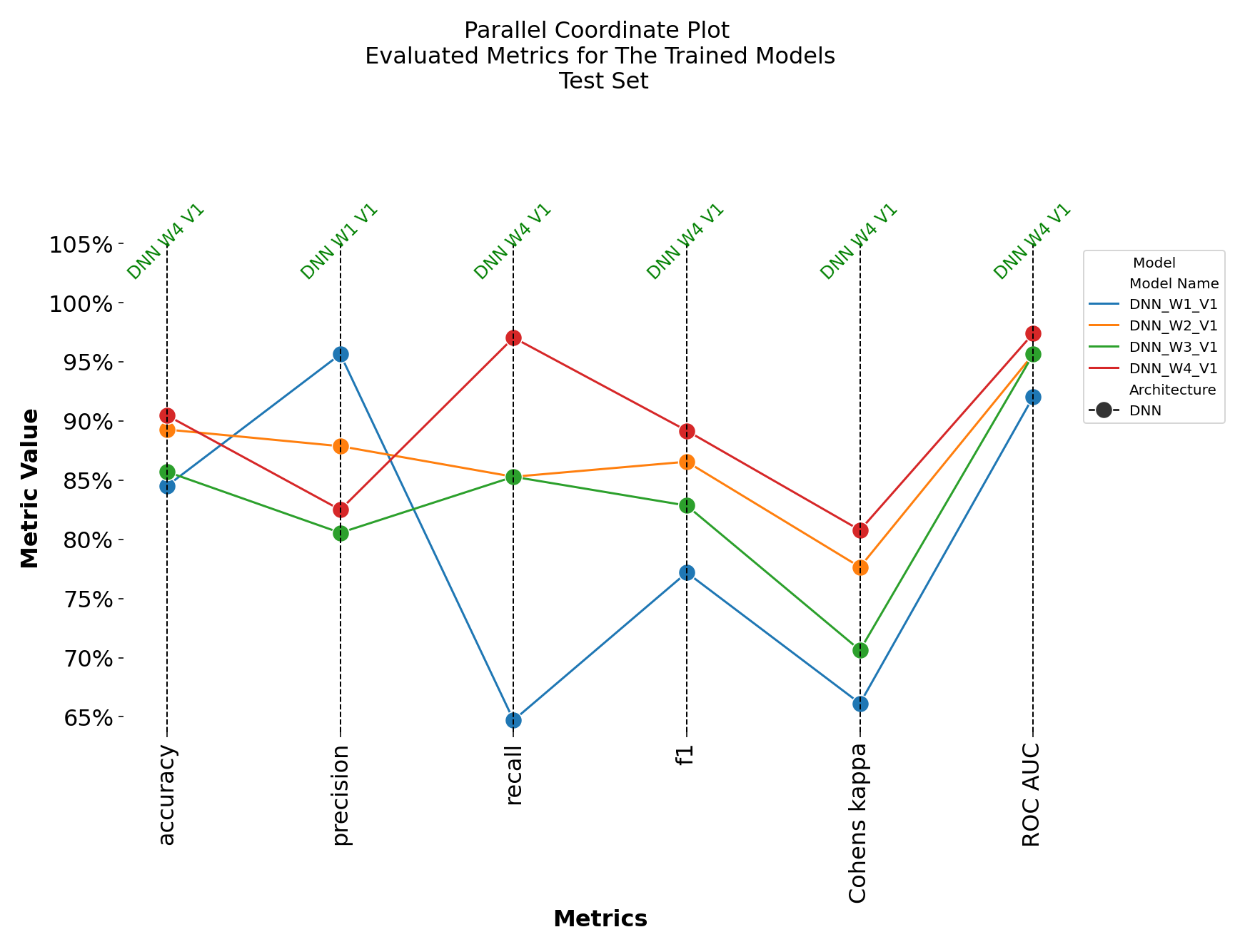}
    \caption{Metrics evaluated in the first group of models. We observe  that DNN-W4-V1 outperforms in most of the metrics.}
    \label{fig:grupo1_test}
\end{figure}

\begin{table}[H]
\centering
\begin{tabular}{|c|c|c|}
\hline
\textbf{Model} & \textbf{Window} & \textbf{Accuracy} \\ \hline
DNN\_W1\_V1 & W1 & 0.84 \\ \hline
DNN\_W2\_V1 & W2 & 0.89 \\ \hline
DNN\_W3\_V1 & W3 & 0.85 \\ \hline
\textbf{DNN\_W4\_V1} & \textbf{W4} & \textbf{0.91} \\ \hline
\end{tabular}%
\caption{Model accuracy comparison for different HU window selections. The best performing model corresponds to window W4.}
\label{tab:hu_window_selection}
\end{table}
Figure~\ref{fig:grupo1_test} is a visual representation of the performance metrics obtained by various models on the test data set.  X-axis denotes the individual performance metrics, and the Y-axis represents the corresponding metric values ranging from 0 to 1. Each line corresponds to a specific model, showing its performance across multiple metrics simultaneously. At the top of each evaluated metric, the name of the model that achieves the highest value for that particular metric is prominently displayed, providing a quick reference for identifying the highest performing models. In addition, an accompanying legend provides information on the architecture type of each model, facilitating comparison and interpretation.

\subsection{Data Augmentation}

After implementing several transformations from the \href{https://github.com/ZFTurbo/volumentations}{volumentations-3D} package, a noticeable decrease in metrics of almost 20 points was observed in comparison with the reference model without augmentation as illustrated in Table.~\ref{tab:modelo_w1_test}. We adopted a rotation transformation in which each 3D CT volume was rotated by a randomly selected angle within a predefined range (-20$^{\circ}$ to +20$^{\circ}$). This rotation helped augment the training data set by introducing variations in the orientation of the CT scans, thus improving the robustness of the model to different imaging perspectives.

\begin{table}[H]
\tiny  
\centering
\begin{tabular}{|c|c|c|c|c|c|c|c|}
\hline
\textbf{Model Name} & \textbf{Dataset} & \textbf{Accuracy} & \textbf{Precision} & \textbf{Recall} & \textbf{F1} & \textbf{Cohen's Kappa} & \textbf{ROC AUC} \\ \hline
DNN\_W1\_V1 & Test & 0.8452 & 0.9565 & 0.6471 & 0.7719 & 0.6613 & 0.9206 \\ \hline
DNN\_W1\_V2 & Test & 0.6429 & 0.6429 & 0.2647 & 0.3750 & 0.1818 & 0.7400 \\ \hline
\end{tabular}%
\caption{Comparison of metrics for the neural network using the W1 window versus its counterpart with data augmentation applied through the Volumentations-3D package. V1 represents the initial model, while V2 corresponds to the model with data augmentation implemented.}
\label{tab:modelo_w1_test}
\end{table}

\subsection{Deterministic Models}

Once the best performing HU window was defined for our case study and making use of the \href{https://github.com/ZFTurbo/classification_models_3D}{classification-models-3D} library, popular neural network architectures were explored in their native form and with minor modifications, such as adding GlobalAveragePooling3D and GlobalMaxPooling3D layers, as well as adjusting the number of filters in the convolutional layers. However, the results of these architectures did not outperform the initial Keras model architecture using the W4 window mentioned in section \ref{r_ventana_hu}. Thus, the following experiments were focused on optimizing the hyperparameters of the deterministic neural network model selected from the initial group (DNN\_W4\_V1). Leveraging the keras-tuner package \cite{omalley2019kerastuner} with the Hyperband tuner, an exhaustive search was performed on various hyperparameter configurations. The Hyperband tuner optimizes the search process by iteratively discarding hyperparameter configurations with poor performance, enabling efficient exploration of the hyperparameter space. The optimal hyperparameters obtained were three blocks comprise of CNNs plus MaxPool and BatchNormalization with 128 neurons each, followed by a Global MaxPool and a dense layer with 256 neurons. The initial learning is 0.001 and a dropout rate of 0.2.  With these hyperparameters, we achieved the best performing deterministic model among the options considered.

\begin{figure}[H]
\centering
\includegraphics[width=1\textwidth]{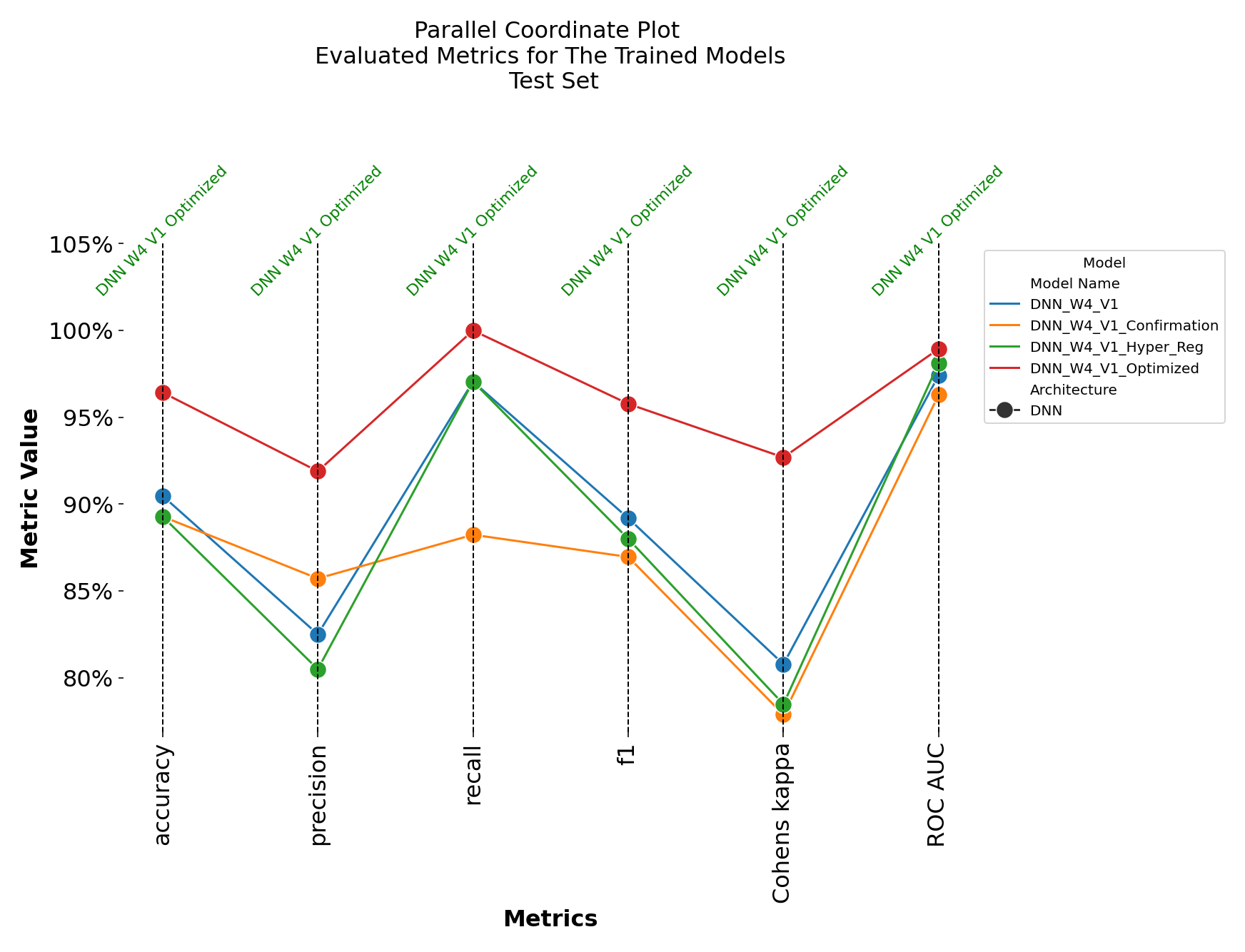}
\caption{Contrast of metrics evaluated in deterministic models before and after optimizing hyper-parameters.}
\label{fig:deterministas_ganadores_test}
\end{figure}

\begin{table}[H]
\tiny  
\centering
\begin{tabular}{|c|c|c|c|c|c|c|c|}
\hline
\textbf{model\_name} & \textbf{Dataset} & \textbf{accuracy} & \textbf{Prec} & \textbf{recall} & \textbf{f1} & \textbf{Cohens kappa} & \textbf{ROC AUC} \\ \hline
DNN\_W4\_V1 & Test & 90\% & 83\% & 97\% & 89\% & 81\% & 97\% \\ \hline
DNN\_W4\_V1\_Optimized & Test & 96\% & 92\% & 100\% & 96\% & 93\% & 99\% \\ \hline
\end{tabular}%
\caption{Comparison of Deterministic Neural Network Metrics for Window W4 and its Optimized Version}
\label{tab:modelo_w4_comparison}
\end{table}
The performance metrics of the DNN\_W4\_V1\_Optimized model listed in Table~\ref{tab:modelo_w4_comparison} showed significant improvement compared to its original version. In particular, the optimized model demonstrated improved classification capabilities across several metrics, particularly evident in the Test dataset. For example, the area under the curve (AUC) increased from 97\% to 99\%, the accuracy increased from 90\% to 96\%, illustrating greater discriminatory power of the model. As a last exercise, regularization techniques were applied to the optimized deterministic model, however, satisfactory results were not achieved as illustrated in Fig.(~\ref{fig:deterministas_ganadores_test}), where the DNN\_W4\_V1\_Hyper\_Reg reduces the performance of the optimized model. As a result, the optimized architecture (DNN\_W4\_V1\_Optimized) was designated as the representative model among its deterministic counterparts.

\begin{figure}[H]
\centering
\includegraphics[width=1\textwidth]{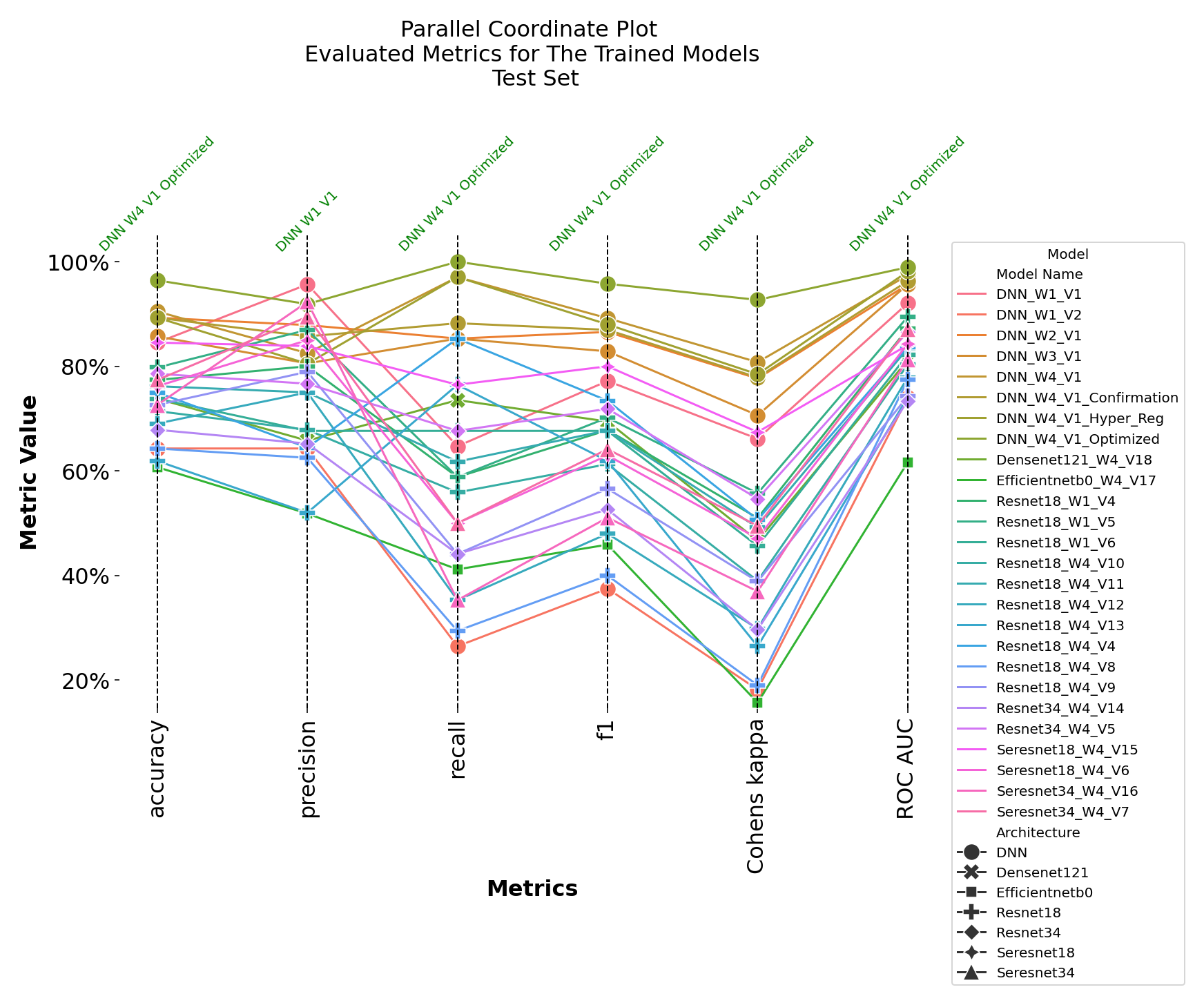}
\caption{Contrast of metrics evaluated in all deterministic models.}
\label{fig:deterministas_test_todos}
\end{figure}

To conclude, the deterministic neural networks explored in this study have provided valuable information on the classification of COVID-19 pneumonia in 3D CT scans. Through rigorous experimentation and optimizations, key architectural configurations and hyperparameters have been identified that contribute to improving model performance. Despite the challenges encountered, such as the limited effectiveness of regularization, the optimized deterministic architecture emerges as a promising solution for accurate classification tasks. As we transition to Bayesian neural networks for further exploration, lessons learned from deterministic models serve as a foundation for future model development and refinement. In the parallel coordinate graph Fig.(~\ref{fig:deterministas_test_todos}), all the results of the deterministic models (more details of each model can be found in \ref{anexo_tabla_modelos_1}, \ref{anexo_tabla_modelos}) evaluated are summarized, providing an overview of its performance in various metrics.
\subsection{Bayesian Neural Networks}

The next exploration step delved into the training of Bayesian neural networks, starting with minor modifications to the architecture of the optimized deterministic model. The main alteration involved replacing the last dense layer of the deterministic model with its Bayesian counterparts. Several layers were explored to evaluate their impact on model performance. These included the integration of  dense Flipout,  Local Reparameterization, and Reparameterization at the top of the network. After exploring the variations in the last layer, the next step involved replacing the deterministic convolutional layers with their Bayesian counterparts in three different scenarios:

\begin{itemize}
    \item Approach 1: 3D MNF Convolutional Layers: In this approach, the deterministic convolutional layers were replaced by 3D Bayesian convolutional layers of multiplicative normalising flows (MNF), leveraging the results of the research in \cite{mnf}.
    
    \item Approach 2: Convolution3DFlipout Convolutional Layers: This approach involved replacing the deterministic convolutional layers with Flipout convolutional layers
    
    \item Approach 3: Convolutional Layers Convolution3DReparameterization: In this approach, the deterministic convolutional layers were replaced with reparametrization convolutional layers.
    
\end{itemize}

\begin{figure}[H]
\centering
\includegraphics[width=1\textwidth]{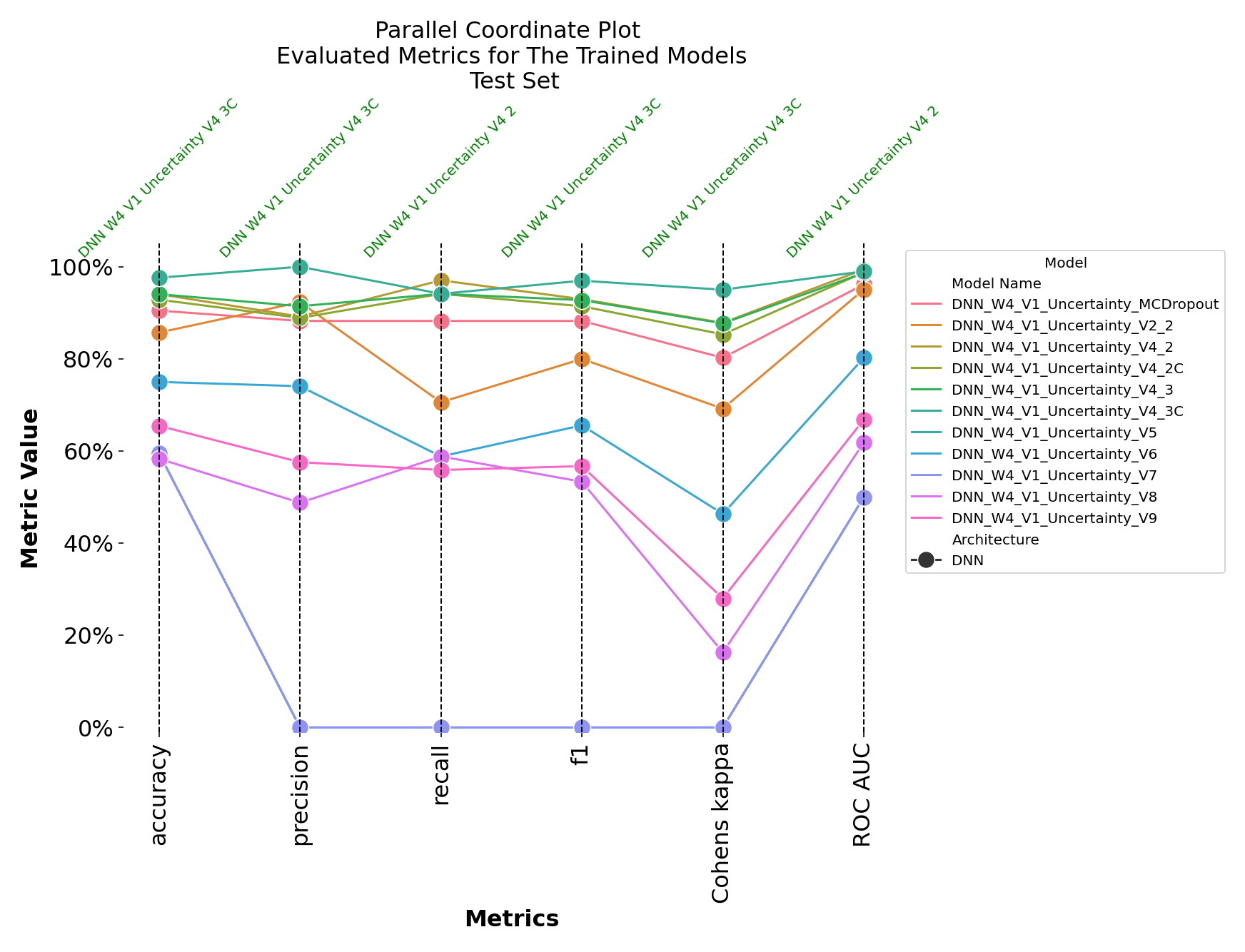}
\caption{Performance of different BNNs methods based on metrics score. We report that MNF technique outperforms the rest of the models.}
\label{fig:bayesianas_test}
\end{figure}

In the modified dropout layer, working not only during training but also in the phase test allows to work under a BNN where the posterior is considered as Gaussian. After evaluating several Bayesian architecture alternatives, the findings in Fig.(~\ref{fig:bayesianas_test}) indicate that the Bayesian neural network comprising multiplicative normalising flows (MNF) convolutional layers and a dense output layer of MNF demonstrates superior performance on the metrics evaluated on the testing data set. This Bayesian architecture achieves comparable performance to the selected deterministic neural network, demonstrating its effectiveness in classifying COVID-19 pneumonia in 3D CT scans. By leveraging the MNF convolutional layers and the dense MNF output layer, this Bayesian neural network effectively captures the underlying uncertainty inherent in the data, thereby improving its predictive capabilities. These results underline the importance of Bayesian approaches in medical imaging tasks, particularly in providing robust and reliable predictions while quantifying uncertainty. The following parallel coordinate plot summarizes the performance of all tested Bayesian models and provides valuable insights into their comparative performance and highlights the effectiveness of the MNF-based Bayesian neural network (Model: DNN\_W4\_V1\_Uncertainty\_V4\_3C). 

\subsection{Calibration} \label{r_calibration_3d}

The calibration analysis of the models involved the visualization of reliability diagrams and the calculation of the Expected Calibration Error (ECE). These analyzes were performed for a variety of models\footnote{The description of each model can be found in \ref{anexo_tabla_modelos}} in order to evaluate their calibration performance and identify optimal configurations.


In addition to evaluating different models, experiments were performed to evaluate the impact of changing the probabilistic Bernoulli layer of the Bayesian models to a \textsl{Sigmoid} activation layer.

The results of the calibration analysis revealed several key insights:

\begin{itemize}
    \item Effect of Sigmoid activation layer: Replacing the probabilistic Bernoulli layer with a Sigmoid activation layer in Bayesian neural networks does not appear to significantly affect the quality of class predictions. Parallel coordinate plot comparisons between models with and without the Sigmoid activation layer illustrated consistent performance on most 
    classification metrics.

    \item Optimal threshold: A notable finding was the identification of an optimal threshold for class 1 predictions. While the default threshold had been set to 0.5 in previous experiments, the analysis revealed better performance of the metrics when the threshold was set to 0.4. This adjustment resulted in higher classification accuracy and other related metrics. 

    \item Reliability and ECE metrics: Among the models evaluated, two of them emerged as the best performers in terms of reliability and ECE metrics. For deterministic models, DNN\_W4\_V1\_Optimized(Fig.(~\ref{fig:conf_determinsita})) consistently demonstrated strong calibration at different threshold settings. His reliability histogram exhibited closely aligned vertical lines and a high level of confidence, indicative of reliable predictions. Similarly, for Bayesian models, DNN\_W4\_V1\_Uncertainty\_V4\_3C (Fig.(\ref{fig:conf_bayesiana})) with the Sigmoid layer and a threshold of 0.4 showed excellent calibration characteristics, with closely grouped vertical lines and minimal deviation from the diagonal. 

\begin{figure}[H]
\centering
\begin{minipage}[t]{0.42\textwidth}
    \centering
    \includegraphics[width=\linewidth]{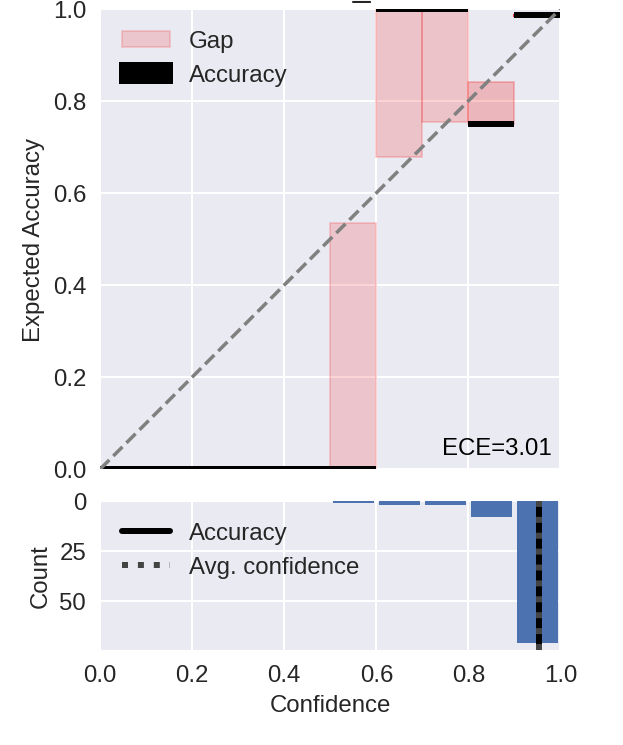}
    \caption{Deterministic Model DNN W4 V1 Optimized - Threshold 0.5}
    \label{fig:conf_determinsita}
\end{minipage}%
\hfill
\begin{minipage}[t]{0.55\textwidth}
    \centering
    \includegraphics[width=\linewidth]{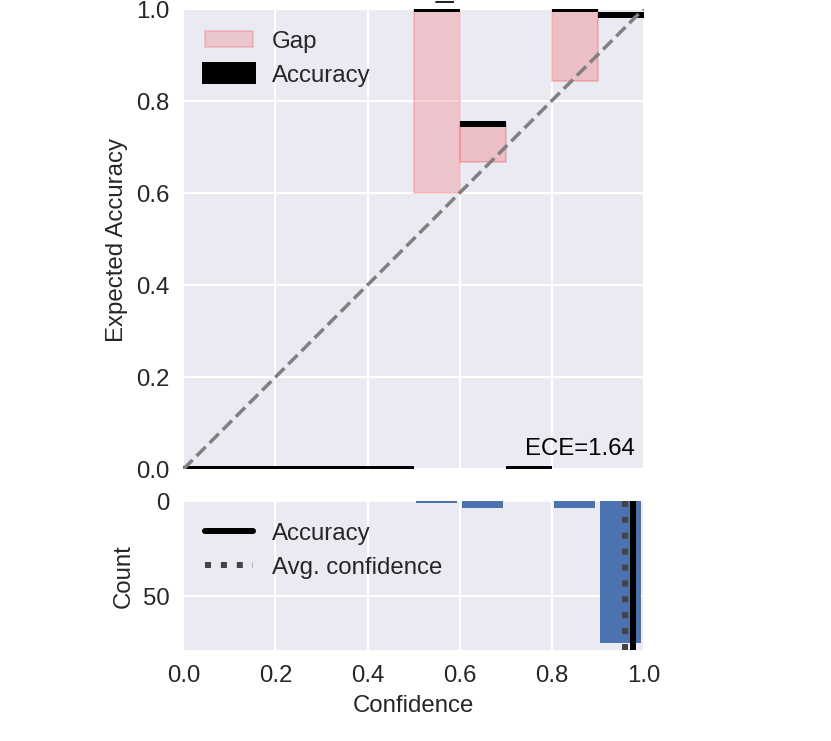}
    \caption{Bayesian Model: DNN W4 V1 Uncertainty V4 3C - Threshold 0.4}
    \label{fig:conf_bayesiana}
\end{minipage}
\caption{Calibration plot showing the relationship between predicted probabilities and observed outcomes for the classification model. A well-calibrated model will have its predicted probabilities closely aligned with the actual outcomes, as represented by the diagonal line - }
\end{figure}



\end{itemize}

The reliability plots in Figs.(\ref{fig:conf_determinsita}, \ref{fig:conf_bayesiana}) of these high-performance models further reinforced the quality of their calibration, and the histograms represent well-calibrated predictions and minimal discrepancies between confidence and precision. These findings underscore the importance of calibration analysis to evaluate the reliability of machine learning models.

\subsection{Uncertainty}

\begin{figure}[H]
    \centering
    \subfloat[CT scan image labeled CT-1. This image is used for model evaluation]{\includegraphics[width=0.6\textwidth]{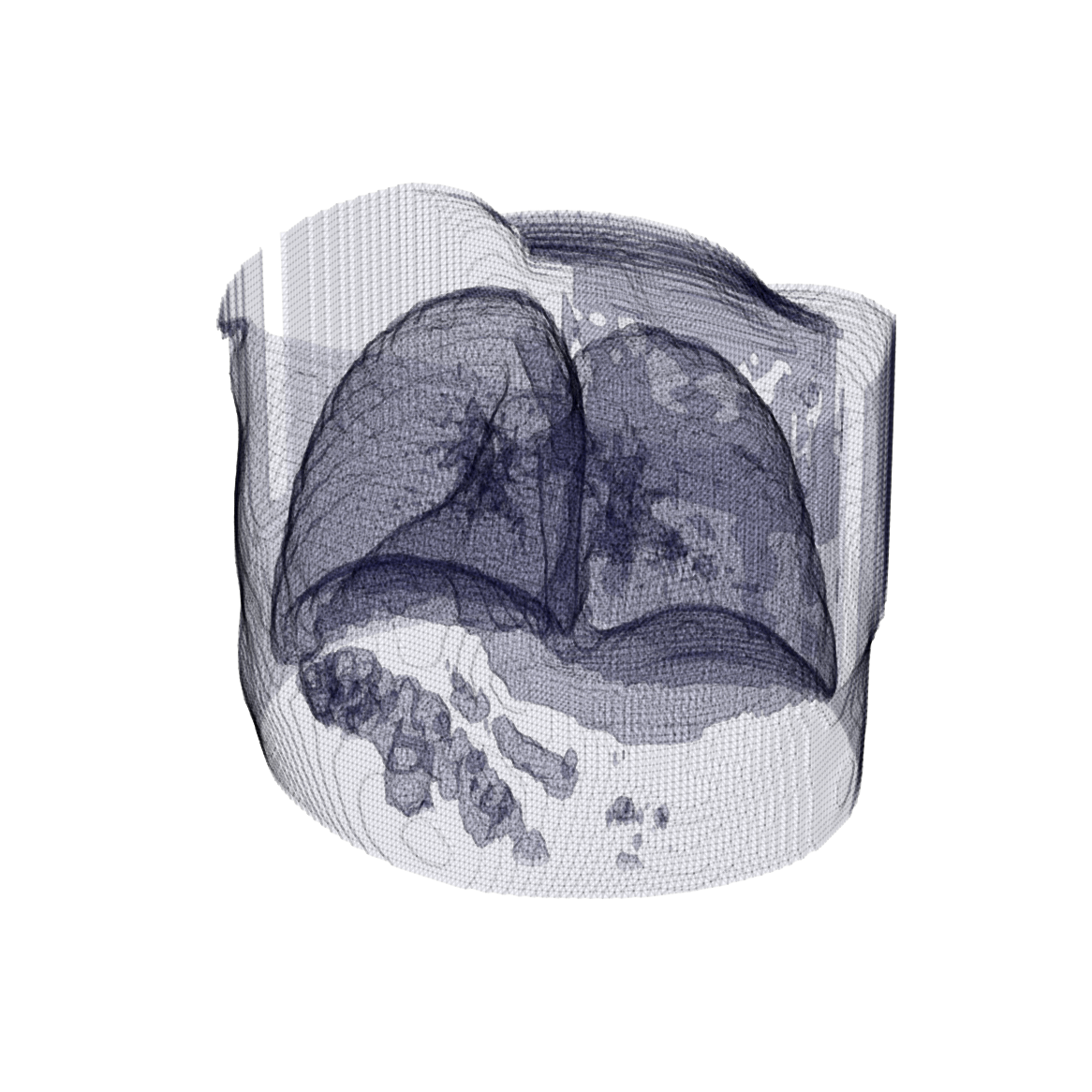}\label{fig:1_ct1_1}}
    \hfill
    \subfloat[Deterministic Model]{\includegraphics[width=0.5\textwidth]{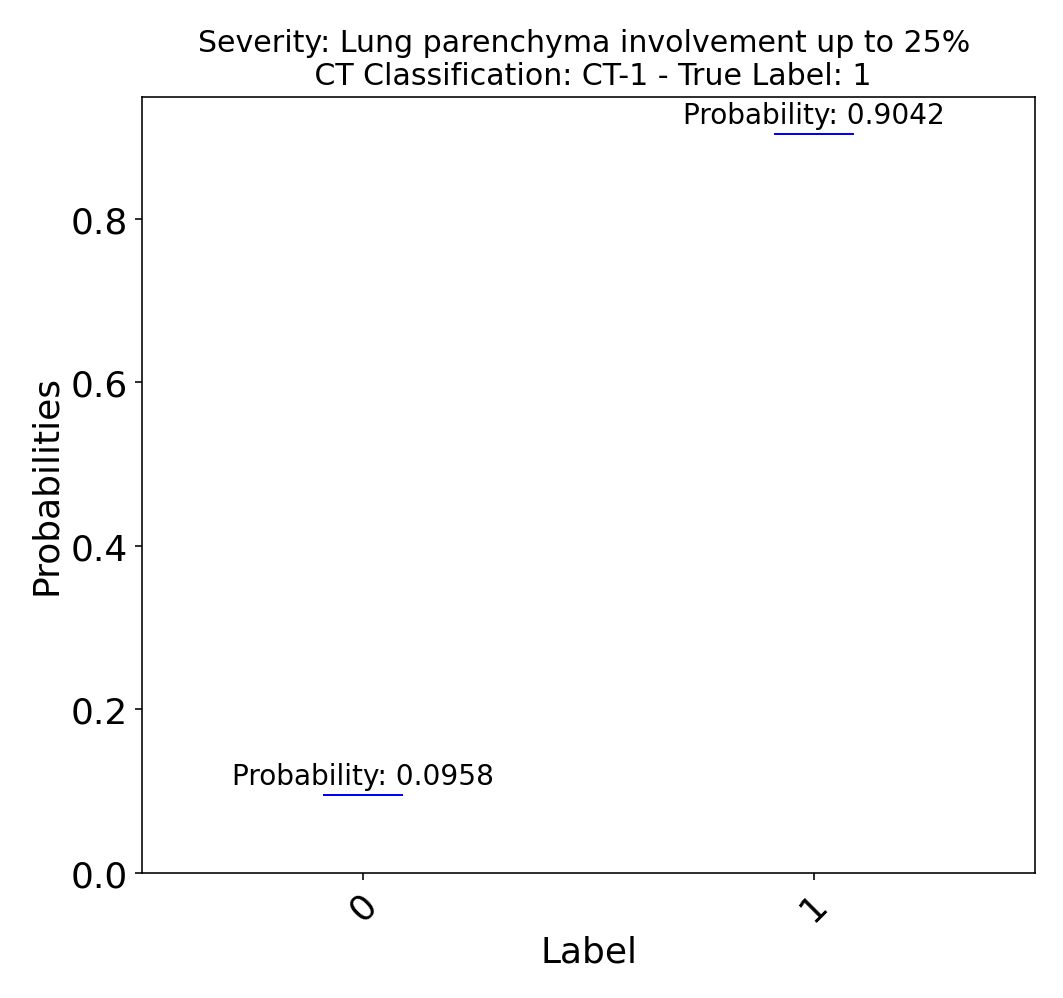}\label{fig:1_ct1_2}}
    \hfill
    \subfloat[Bayesian Model]{\includegraphics[width=0.5\textwidth]{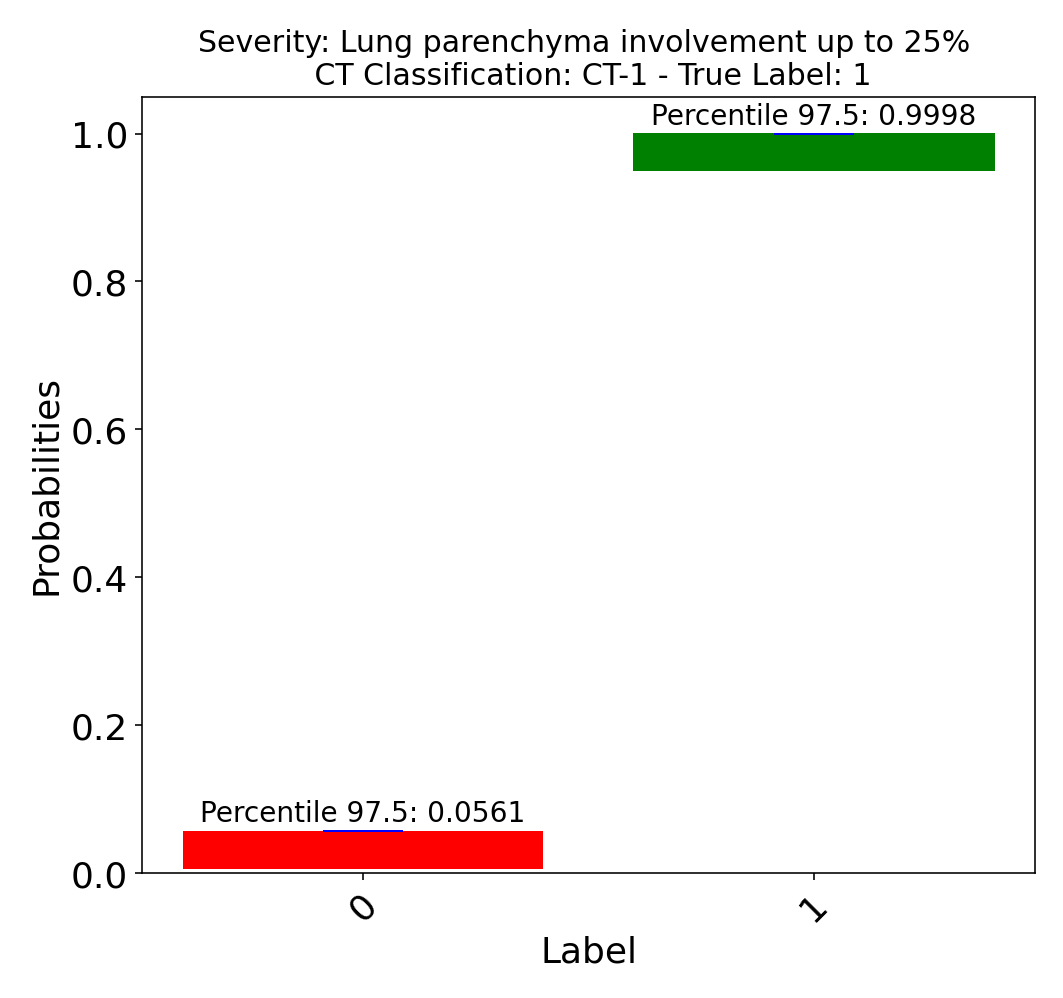}\label{fig:1_ct1_3}}
    \caption{95\% confidence intervals for estimated model probabilities for a deterministic and a Bayesian model, respectively. These plots show the uncertainty in predictions for a CT scan image, where the deterministic model provides fixed probability estimates, while the Bayesian model incorporates uncertainty, offering a range of possible outcomes through the confidence intervals.}
    \label{fig:layout}
\end{figure}

The uncertainty analysis of the models involved the analysis of prediction intervals derived from simulation exercises performed on selected CT images mentioned in the \ref{sec6:metho} section. These images spanned a variety of classes and were used to evaluate the predictive performance and uncertainty estimation capabilities of deterministic and Bayesian models. Six images were randomly selected for analysis, consisting of four images that represent classes with different degrees of pneumonia involvement (CT-1 and CT-4) unknown to the model and two images from the testing data set labeled as not affected by pneumonia (CT-0). The analysis focused on contrasting the predictions and uncertainty (in the Bayesian scenario) of the two best models chosen in previous stages: the deterministic model (DNN\_W4\_V1\_Optimized) and the Bayesian model (DNN\_W4\_V1\_Uncertainty\_V4\_3C +  Sigmoid). These models were selected based on their performance and representativeness in the context of the study.\\
\textbf{CT-1 class images}: In Figure~\ref{fig:layout}, both the deterministic and Bayesian models correctly identified deficiencies in the tomography, classifying it as class 1 (presence of impairment). However, the Bayesian model stood out by offering narrow prediction intervals, indicating high confidence in its predictions. This ability to accurately quantify uncertainty is essential in medical applications, as it allows for more informed and accurate decision making. These results highlight the usefulness of Bayesian models in medical diagnosis. The Bayesian model provides confidence intervals for each class as follows: for Class 0, the interval ranges from $[0.00018005, 0.05612653]$, indicating a low probability, while for Class 1, the interval is much higher, ranging from $[0.94387347, 0.99981995]$, reflecting a strong likelihood for this class. These intervals capture the model's uncertainty in assigning probabilities to each class.

\begin{figure}[H]
    \centering
    \subfloat[CT scan image labeled CT-1. This image is used for model evaluation]{\includegraphics[width=0.6\textwidth]{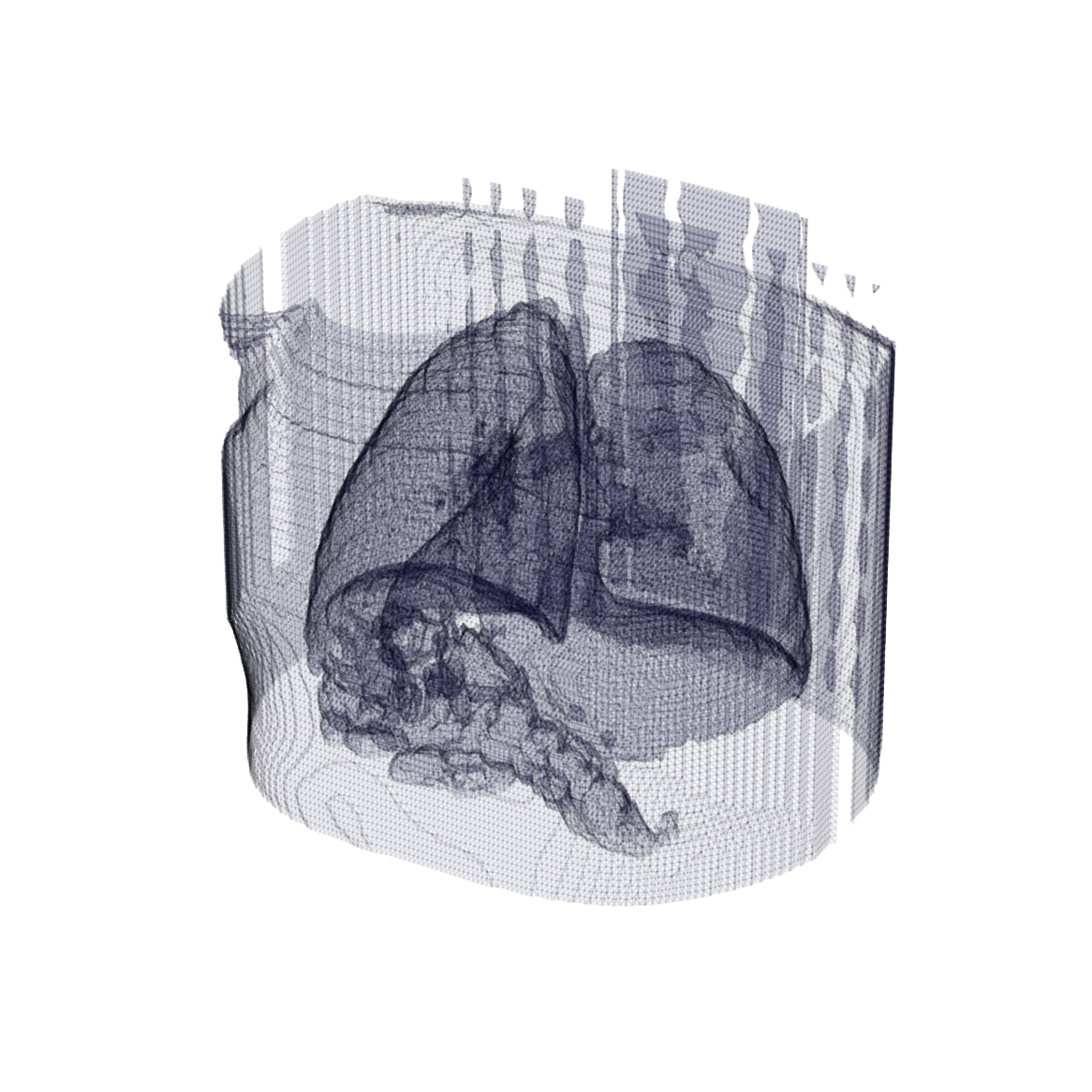}\label{fig:2_ct1_1}}
    \medskip
    
    \subfloat[Deterministic Model]{\includegraphics[width=0.45\textwidth]{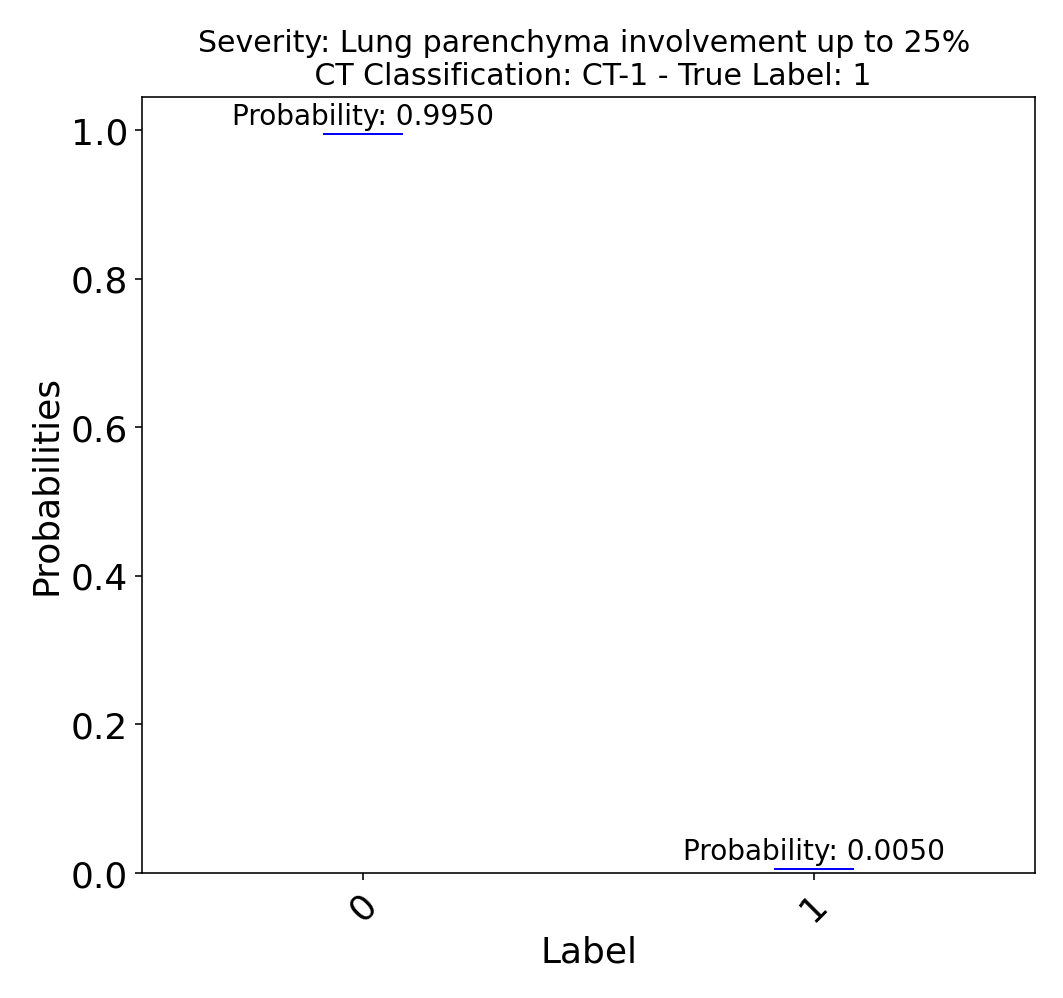}\label{fig:2_ct1_2}}
    \hfill
    \subfloat[Bayesian Model]{\includegraphics[width=0.45\textwidth]{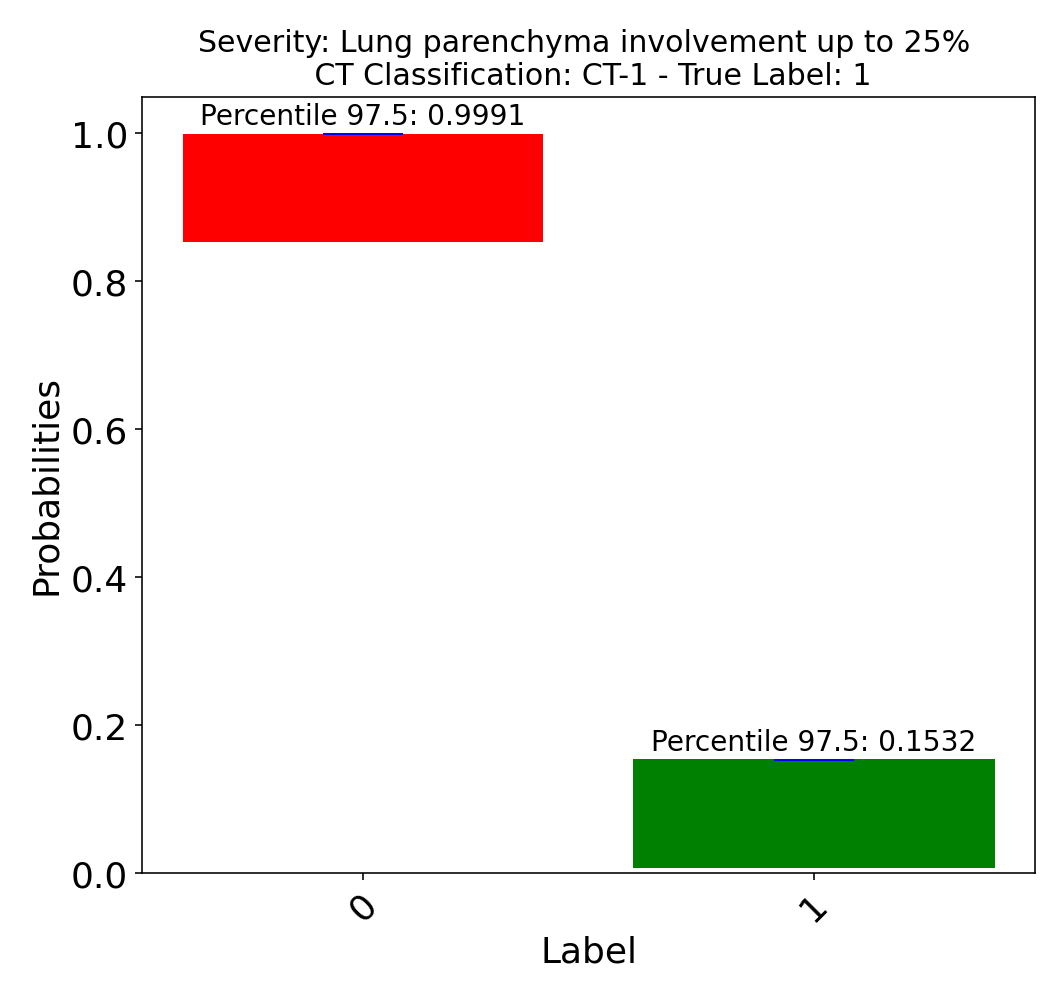}\label{fig:2_ct1_3}}
    
    \caption{95\% confidence intervals for estimated model probabilities for a deterministic and a Bayesian model, respectively. These plots show the uncertainty in predictions for a CT scan image, where the deterministic model provides fixed probability estimates, while the Bayesian model incorporates uncertainty, offering a range of possible outcomes through the confidence intervals.}
    \label{fig:layout_2}
\end{figure}

On the other hand, Fig.~\ref{fig:layout_2} shows that both models agree that the analyzed tomography does not present any involvement (that is, Class 0). However this statement is not precise, since the image belongs to a class with a lung parenchyma involvement of 25\%. Despite this prediction error, BNNs offers remarkably wide confidence intervals, suggesting the presence of uncertainty. This result highlights the importance of considering uncertainty in predictions, even when they are incorrect, thus providing a more cautious evaluation of the medical diagnosis. The BNNs confidence intervals for each class are as follows: for Class 0, the interval ranges from $[0.84678476, 0.99909177]$, suggesting a high probability, while for Class 1, the interval spans from $[0.00090823, 0.15321524]$, indicating a lower likelihood. These intervals reflect the model's certainty in classifying the data between the two classes.\\

\begin{figure}[H]
    \centering
    \subfloat[CT scan image labeled CT-4. This image is used for model evaluation]{\includegraphics[width=0.6\textwidth]{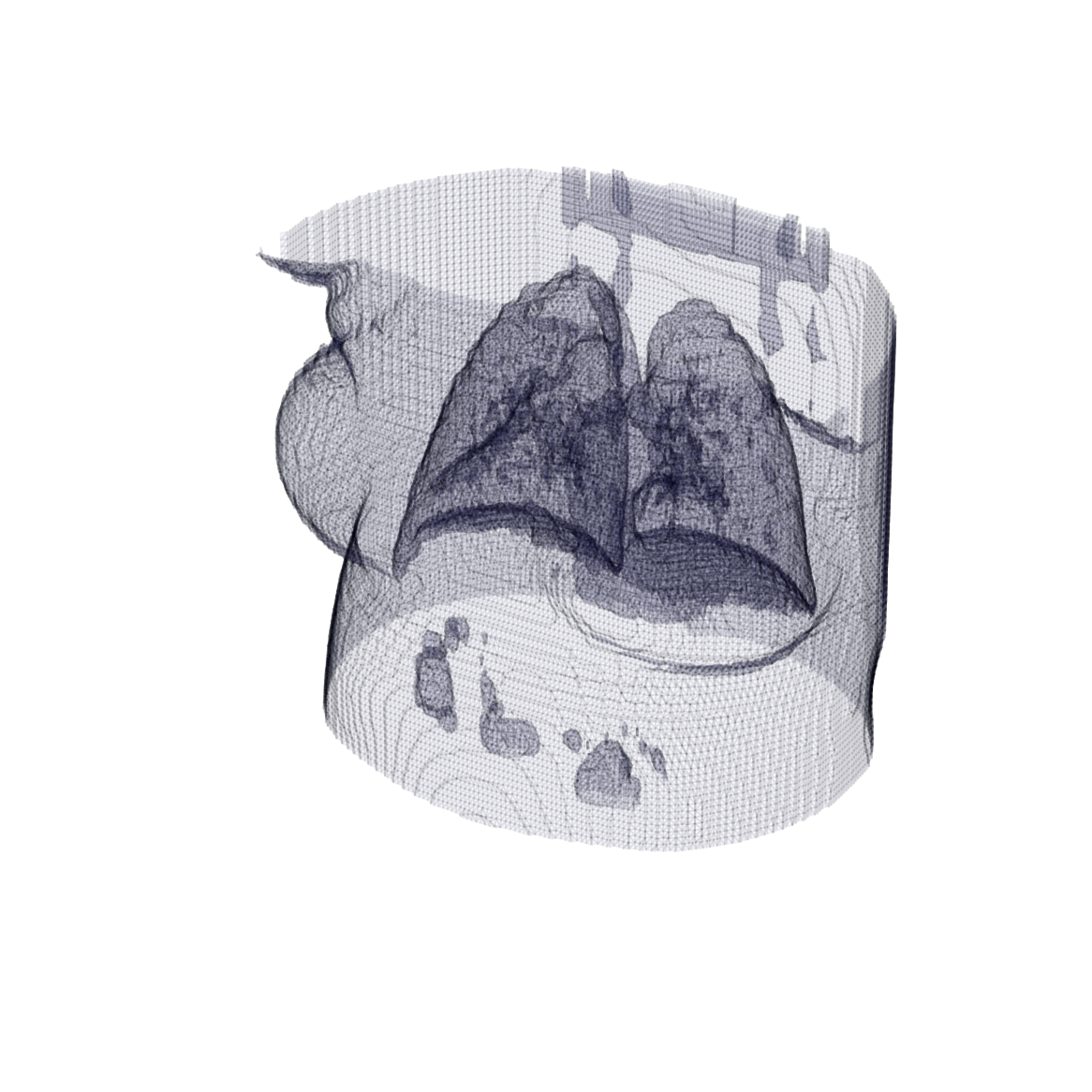}\label{fig:3_ct4_1}}
    
    \medskip
    
    \subfloat[Deterministic Model]{\includegraphics[width=0.45\textwidth]{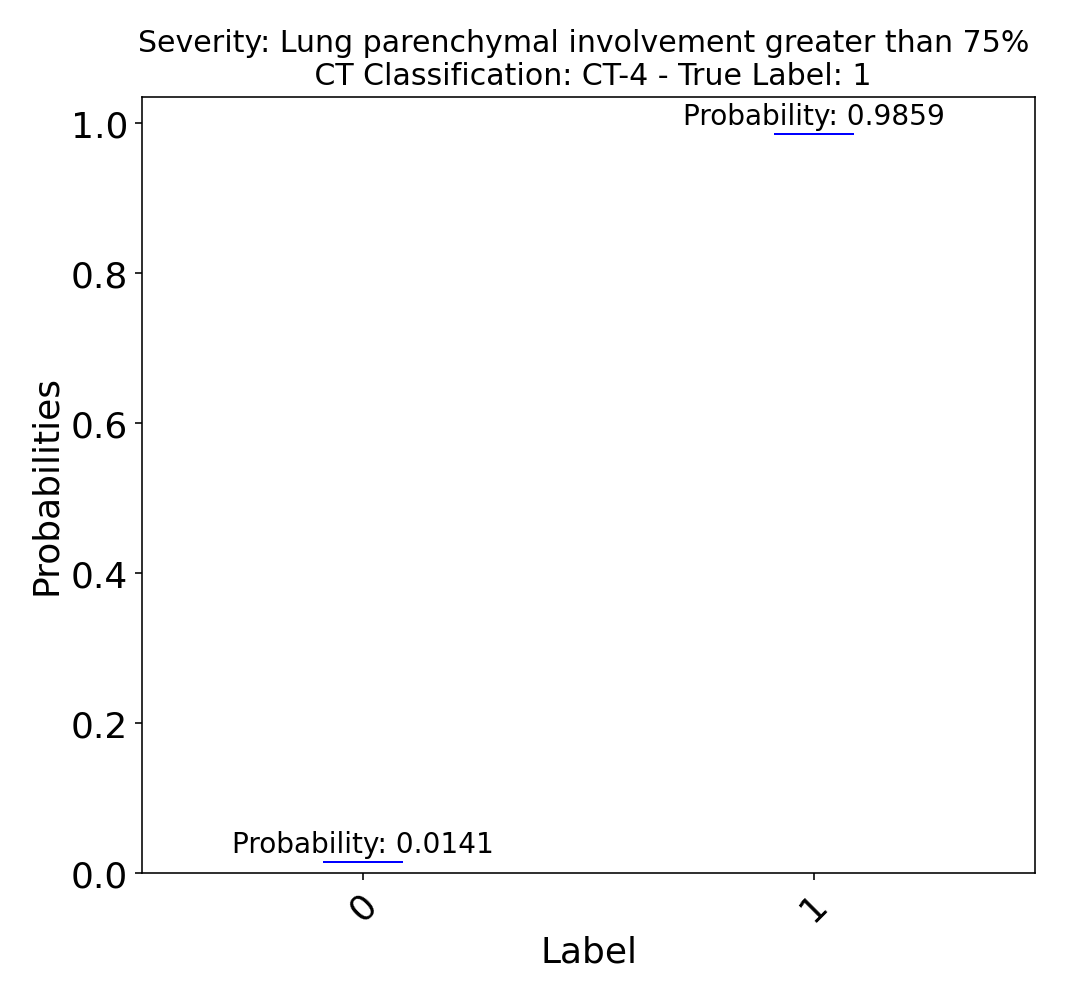}\label{fig:3_ct4_2}}
    \hfill
    \subfloat[Bayesian Model]{\includegraphics[width=0.45\textwidth]{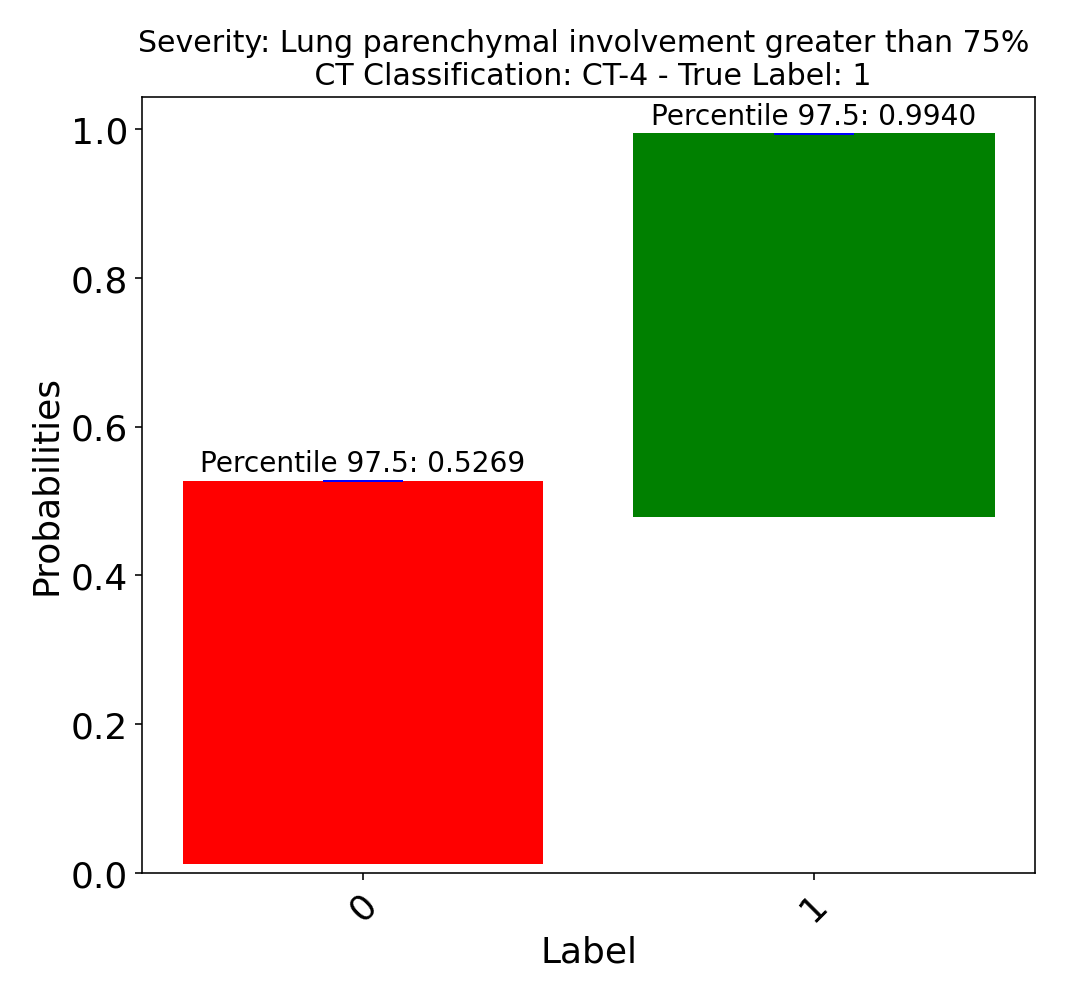}\label{fig:3_ct4_3}}
    
    \caption{95\% confidence intervals for estimated model probabilities for a deterministic and a Bayesian model, respectively. These plots show the uncertainty in predictions for a CT scan image, where the deterministic model provides fixed probability estimates, while the Bayesian model incorporates uncertainty, offering a range of possible outcomes through the confidence intervals.}
    \label{fig:layout_3}
\end{figure}

\textbf{CT-4 class images} In Fig.~\ref{fig:layout_3},  the models correctly identified alterations in the analyzed tomography, classifying it as belonging to Class 1. However, the prediction intervals of the Bayesian model have a fairly considerable width, indicating greater uncertainty in their predictions. This wide uncertainty could be considered a signal to generate alerts or request additional review by medical professionals, in order to ensure a more accurate evaluation and avoid possible diagnostic errors. The confidence intervals provided by the Bayesian model for each class are: for Class 0, the interval ranges from $[0.0060012, 0.52692841]$, reflecting a relatively lower probability. In contrast, for Class 1, the interval spans from $[0.47307159 , 0.9939988]$, indicating a much higher likelihood. These intervals demonstrate the model distribution of certainty between the two classes.

\begin{figure}[H]
    \centering
    \subfloat[CT scan image labeled CT-4. This image is used for model evaluation]{\includegraphics[width=0.6\textwidth]{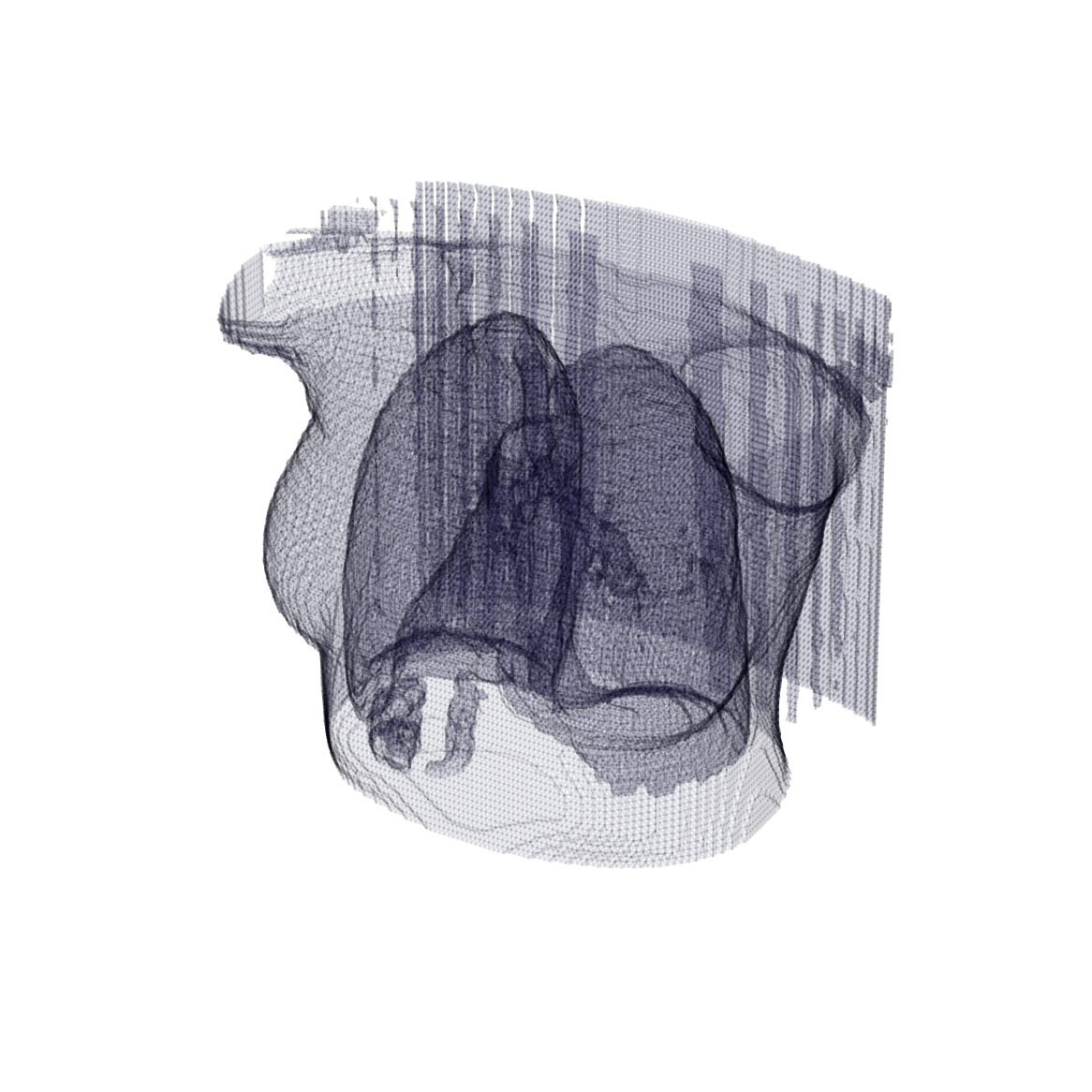}\label{fig:4_ct4_1}}
    
    \medskip
    
    \subfloat[Deterministic Model]{\includegraphics[width=0.45\textwidth]{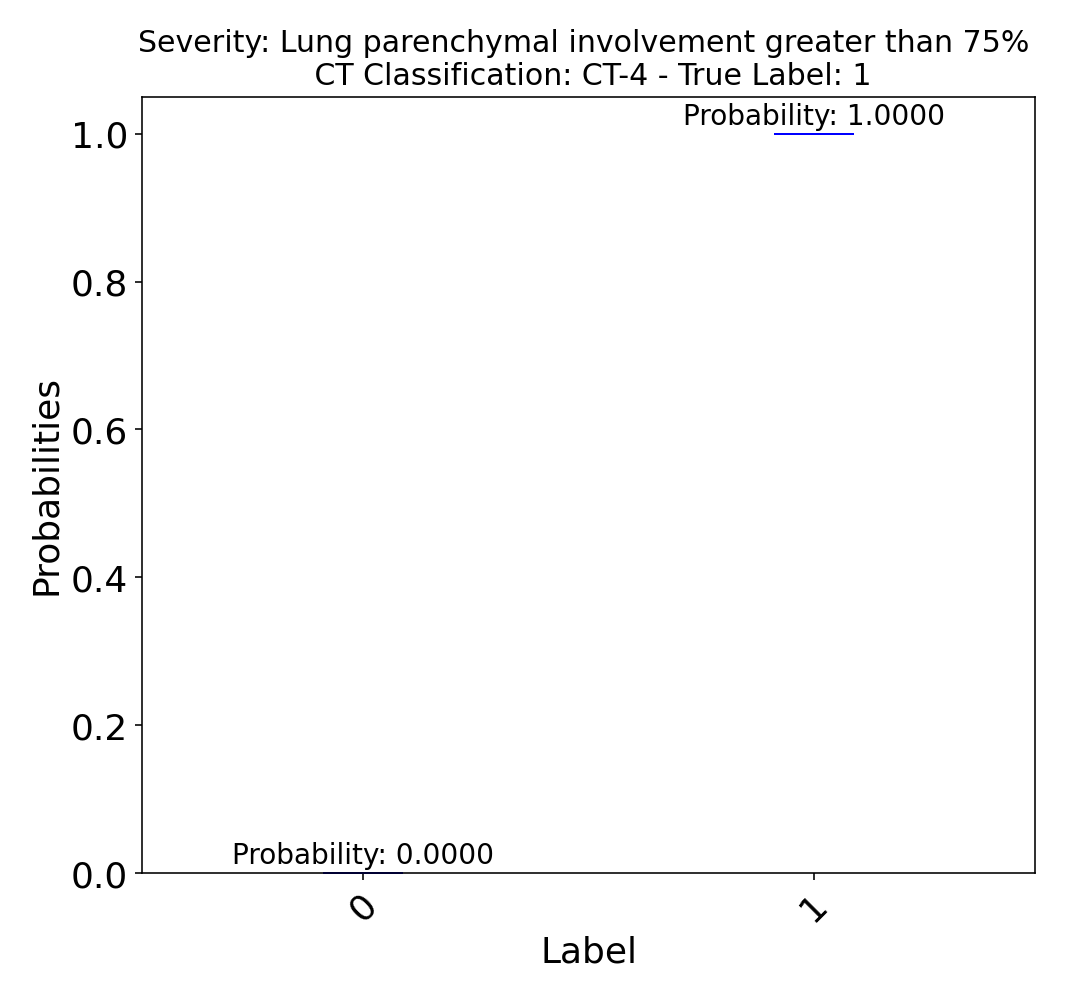}\label{fig:4_ct4_2}}
    \hfill
    \subfloat[Bayesian Model]{\includegraphics[width=0.45\textwidth]{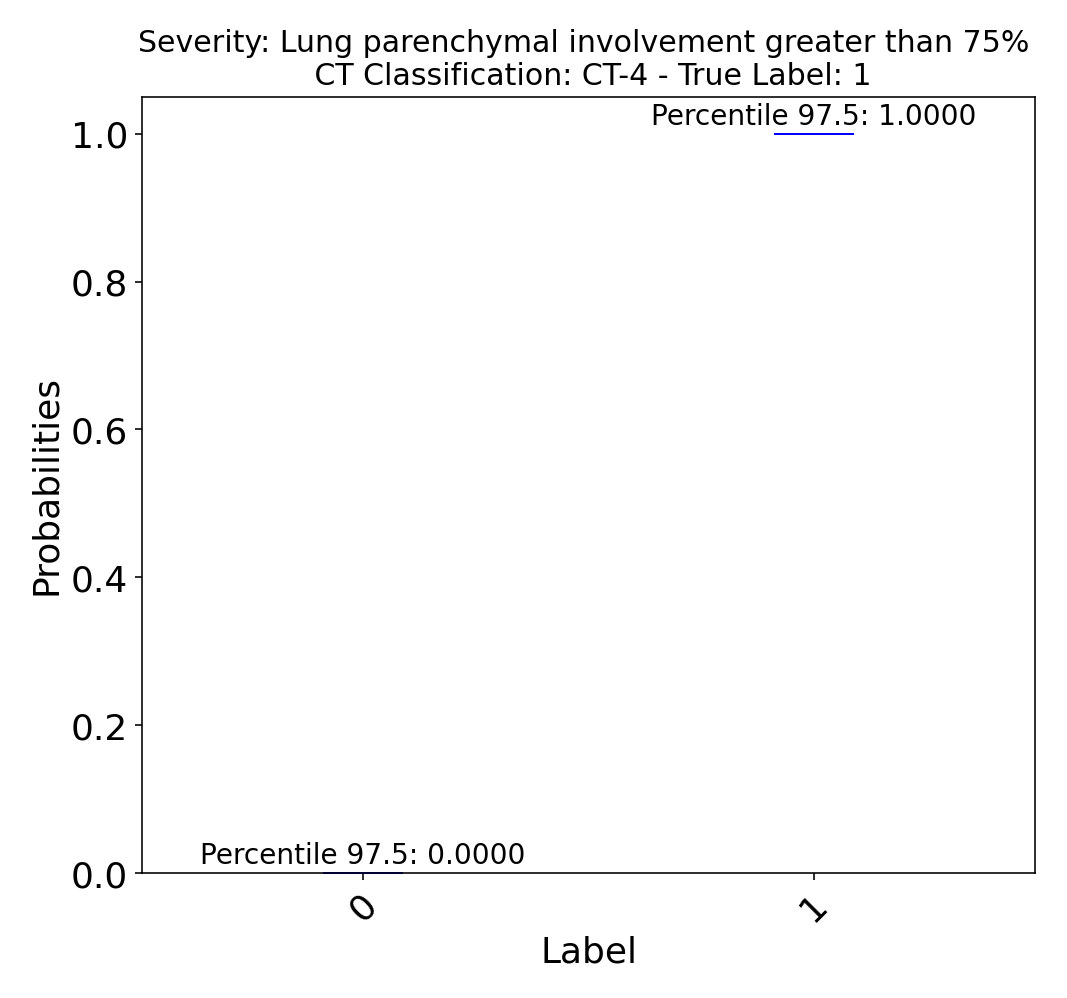}\label{fig:4_ct4_3}}
    
    \caption{95\% confidence intervals for estimated model probabilities for a deterministic and a Bayesian model, respectively. These plots show the uncertainty in predictions for a CT scan image, where the deterministic model provides fixed probability estimates, while the Bayesian model incorporates uncertainty, offering a range of possible outcomes through the confidence intervals.}
    \label{fig:layout_4}
\end{figure}

On the other hand, Fig.~\ref{fig:layout_4} shows that both models correctly identified the deficiencies in tomography as Class 1. However, it is interesting to note that the prediction intervals of the Bayesian model were remarkably narrow, suggesting extremely low uncertainty in its predictions. This narrowness in the prediction intervals of the Bayesian model points to exceptional reliability in its results, highlighting its ability to make accurate and reliable predictions even in complex situations. The Bayesian model's confidence intervals for each class are as follows: for Class 0, the interval is extremely narrow, ranging from $[0, 0.00002287]$, indicating almost no probability for this class. For Class 1, the interval spans from $[0.99997713, 1.0]$, reflecting a near-certain prediction. These intervals clearly illustrate the model's high confidence in favor of Class 1.\\

\begin{figure}[H]
    \centering
    \subfloat[CT scan image labeled CT-0. This image is used for model evaluation]{\includegraphics[width=0.6\textwidth]{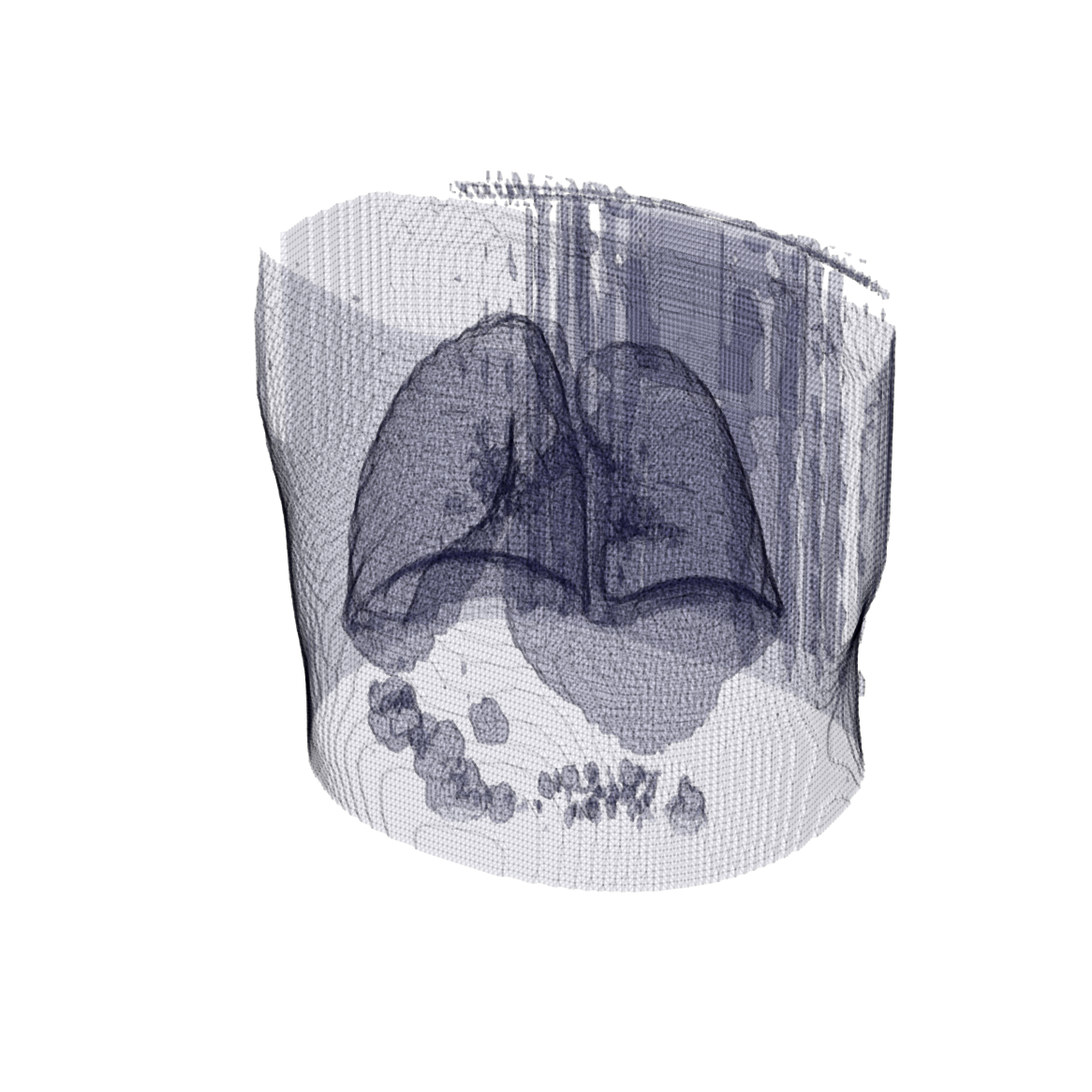}\label{fig:5_ct0_1}}
    
    \medskip
    
    \subfloat[Deterministic Model]{\includegraphics[width=0.45\textwidth]{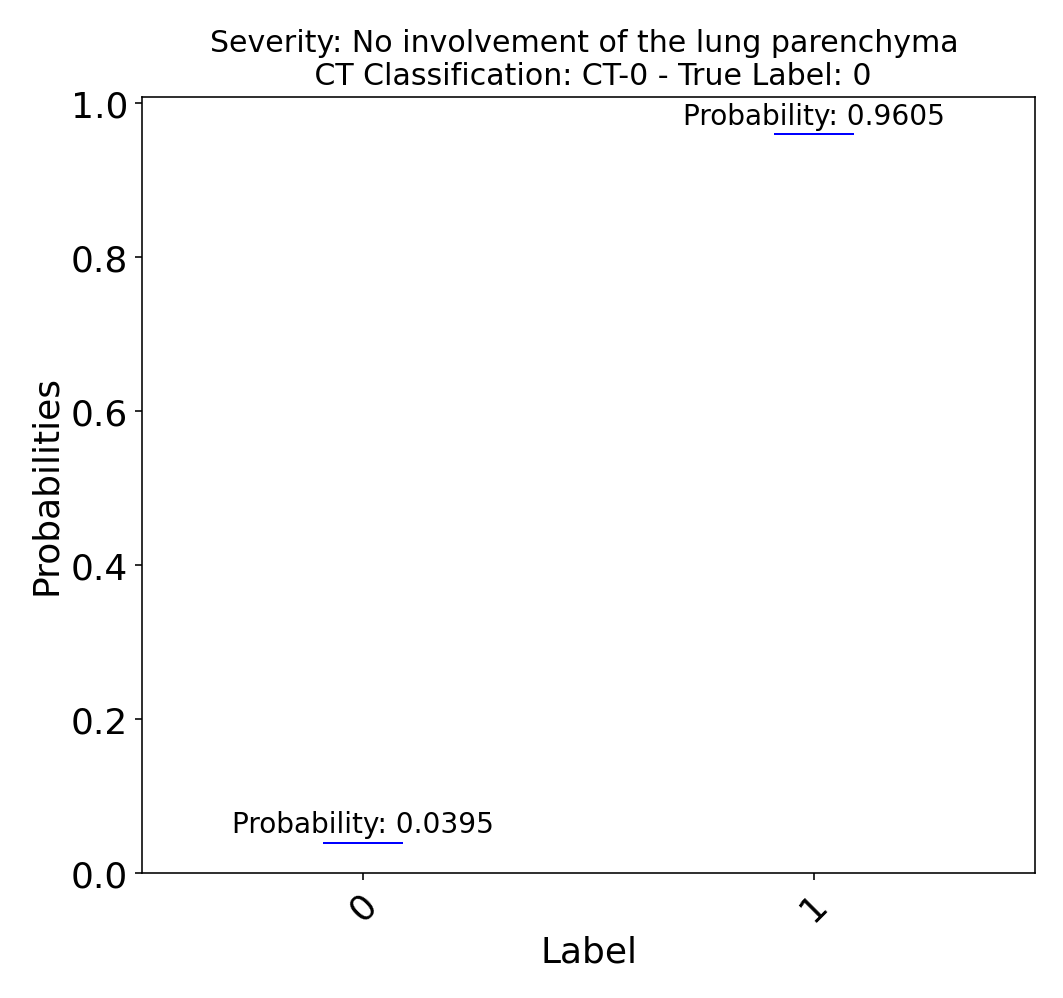}\label{fig:5_ct0_2}}
    \hfill
    \subfloat[Bayesian Model]{\includegraphics[width=0.45\textwidth]{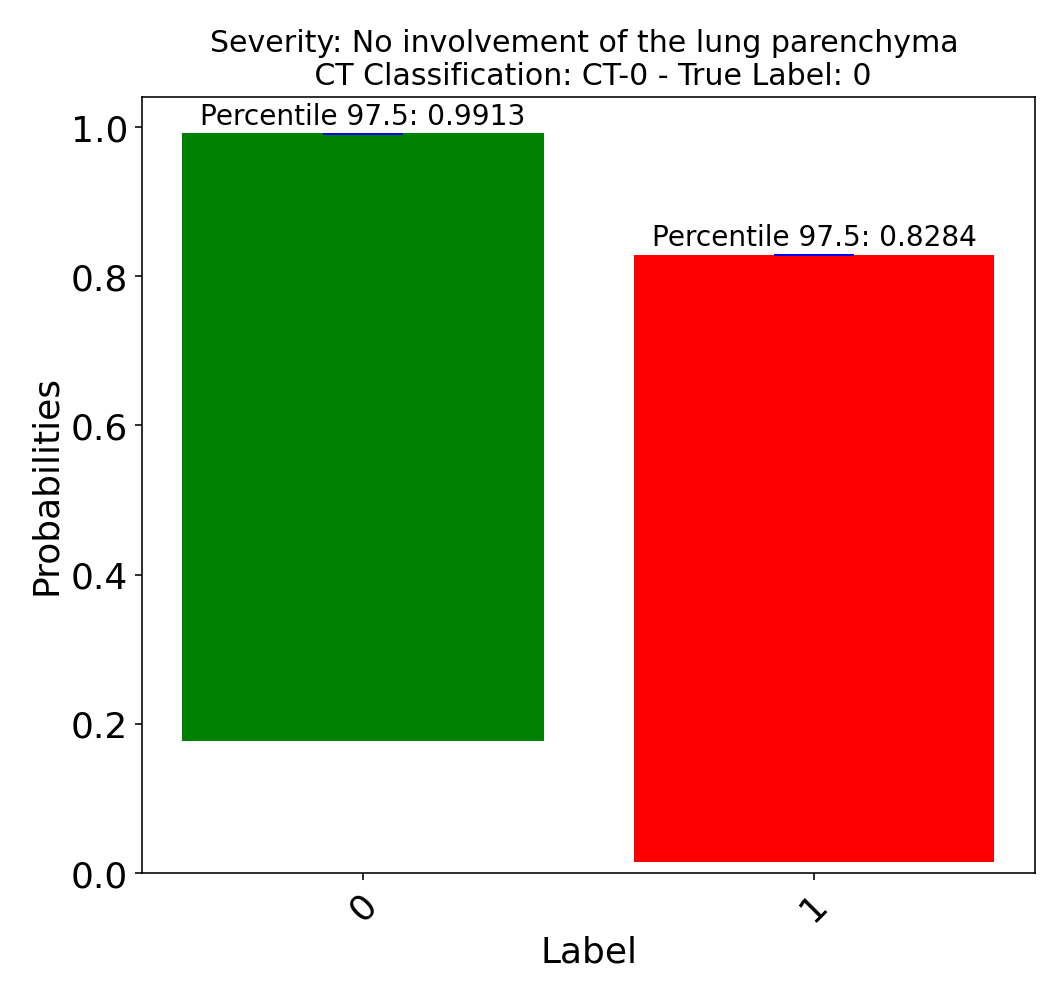}\label{fig:5_ct0_3}}
    
    \caption{95\% confidence intervals for estimated model probabilities for a deterministic and a Bayesian model, respectively. These plots show the uncertainty in predictions for a CT scan image, where the deterministic model provides fixed probability estimates, while the Bayesian model incorporates uncertainty, offering a range of possible outcomes through the confidence intervals.}
    \label{fig:layout_5}
\end{figure}

\begin{figure}[H]
    \centering
    \subfloat[CT scan image labeled CT-0. This image is used for model evaluation]{\includegraphics[width=0.6\textwidth]{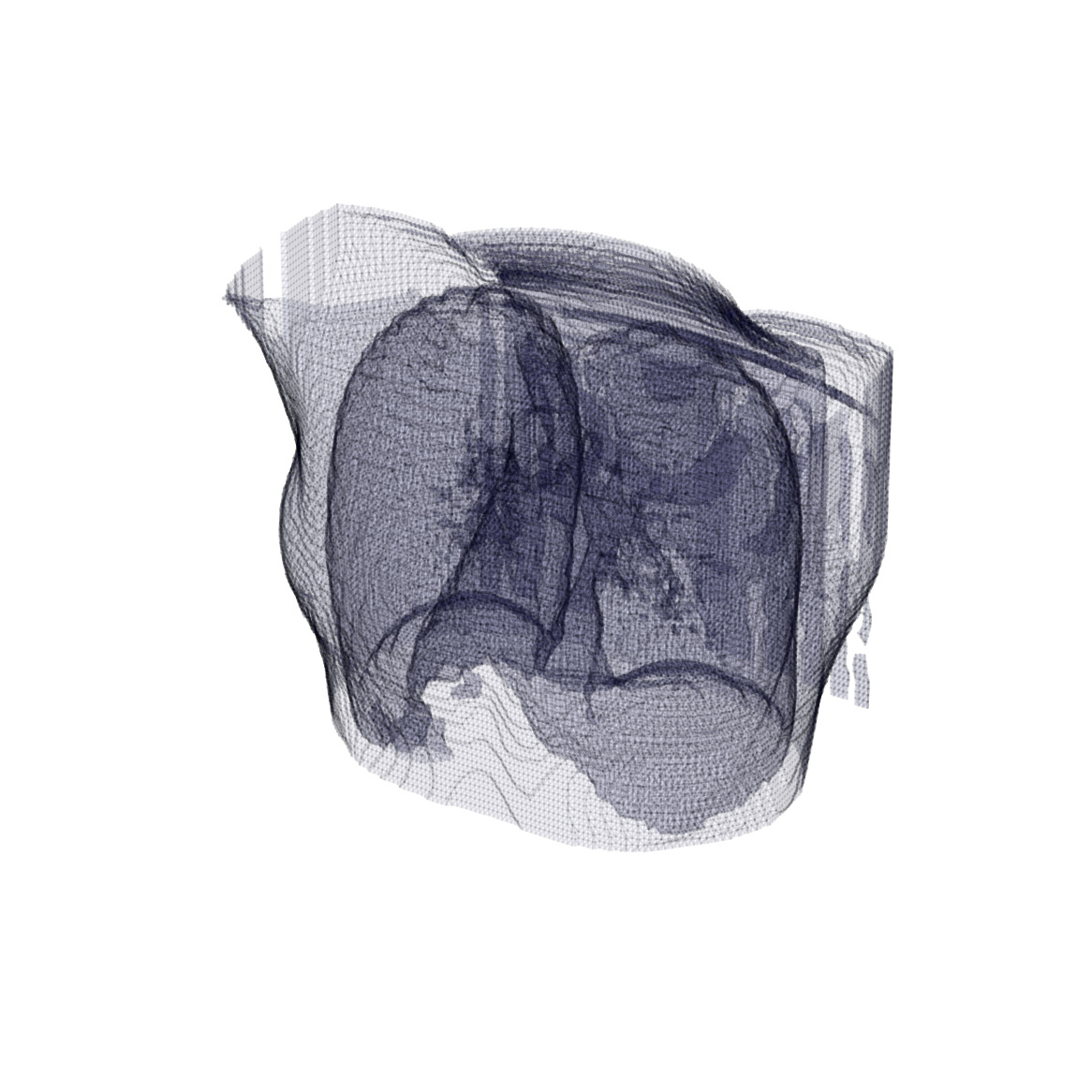}\label{fig:6_ct0_1}}
    
    \medskip
    
    \subfloat[Deterministic Model]{\includegraphics[width=0.45\textwidth]{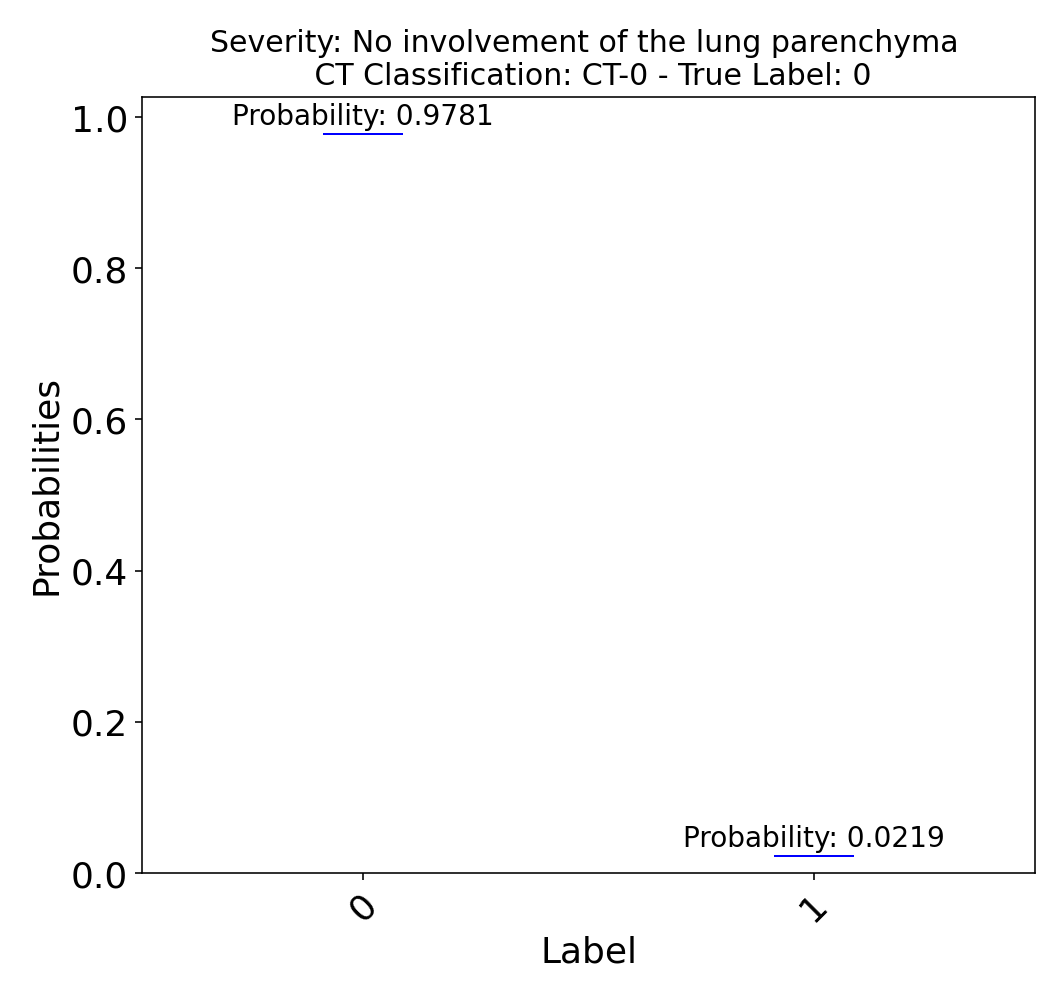}\label{fig:6_ct0_2}}
    \hfill
    \subfloat[Bayesian Model]{\includegraphics[width=0.45\textwidth]{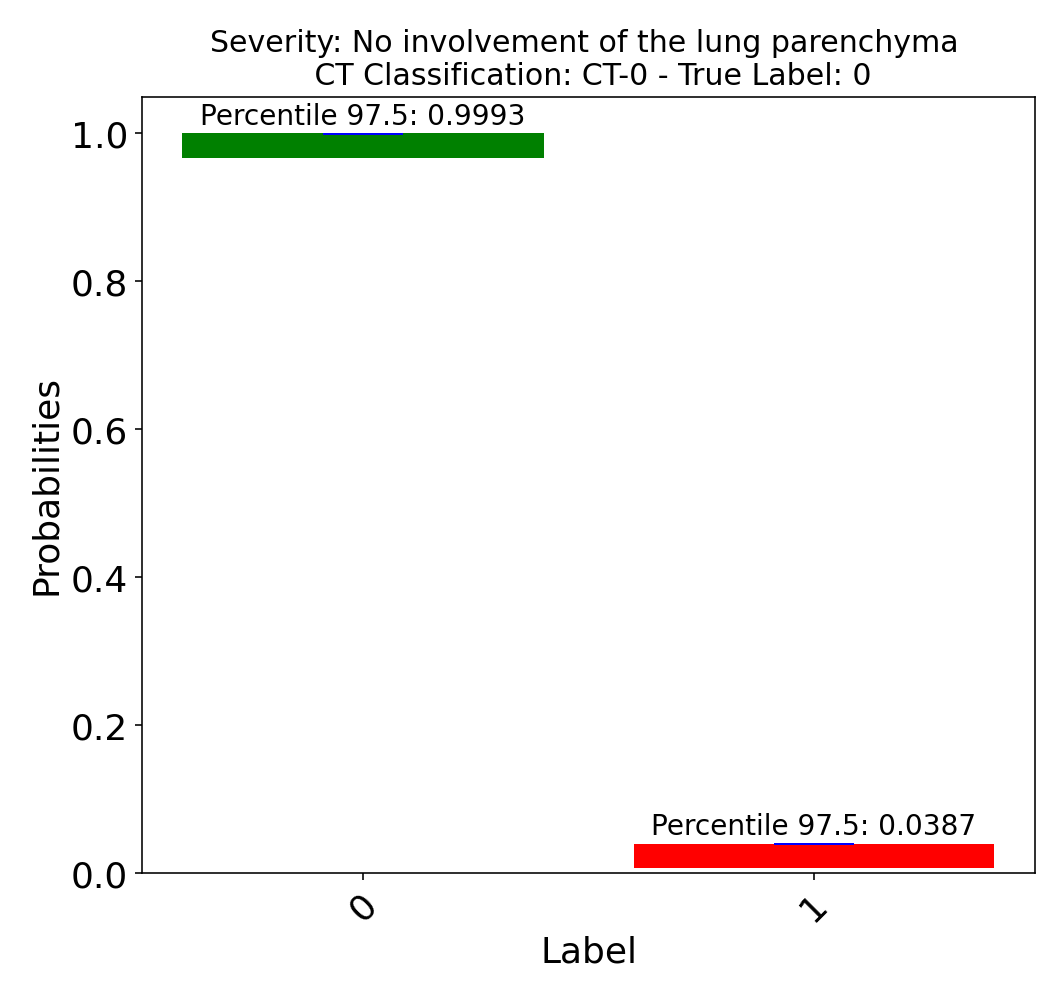}\label{fig:6_ct0_3}}
    
    \caption{95\% confidence intervals for estimated model probabilities for a deterministic and a Bayesian model, respectively. These plots show the uncertainty in predictions for a CT scan image, where the deterministic model provides fixed probability estimates, while the Bayesian model incorporates uncertainty, offering a range of possible outcomes through the confidence intervals.}
    \label{fig:layout_6}
\end{figure}

\textbf{CT-0 class images} In Fig.~\ref{fig:layout_5}, we observe that the deterministic model exhibits a bias towards label 1, while the Bayesian model was biased towards label 0. Even though the actual label was 0 ( absence of anomalies), the wide prediction intervals of the Bayesian model indicate substantial uncertainty. The presence of these wide prediction intervals in the Bayesian model suggests that the model is not completely confident in its prediction, raising an alert for a more detailed clinical diagnosis. This highlights the importance of Bayesian neural networks in the field of medicine as they provide crucial information about the uncertainty associated with predictions, which can help medical professionals make more informed and accurate decisions. The Bayesian model's confidence intervals for each class are as follows: for Class 0, the interval ranges from $[0.17163233, 0.99132968]$, showing a wide range of probabilities and indicating a higher likelihood for this class. For Class 1, the interval spans from $[0.00867032, 0.82836767]$, reflecting a lower but still notable probability. This distribution suggests some uncertainty, with the model leaning more towards Class 0.
Furthermore, Fig.~\ref{fig:layout_6} depicts  that both models accurately identify the absence of anomalies in the analyzed tomography, classifying it as class 0. However, it should be noted that the Bayesian model shows narrow confidence intervals , an indicator of low uncertainty in its predictions. This characteristic of the Bayesian model suggests greater reliability in its results. The Bayesian model's confidence intervals for each class are as follows: for Class 0, the interval is quite narrow, ranging from $[0.96128352 , 0.99934633]$, suggesting a strong likelihood for this class. For Class 1, the interval is from 0.00065367 to 0.03871648, indicating a very low probability. This distribution reflects the model’s high confidence in favor of Class 0.

To conclude, this analysis highlights the importance of estimating uncertainty in model predictions. Despite the model's inaccuracies, the wide prediction intervals provided by the Bayesian network in some images highlighted the presence of uncertainty in the predictions. This recognition of uncertainty is crucial to decision-making processes, as it alerts stakeholders to the model's ''doubt'' about certain predictions and encourages cautious interpretation of results. Importantly, exceptionally wide intervals in Bayesian network predictions can serve as a signal to raise red flags and promote more detailed evaluation by medical professionals.
\subsection{Explainability with SHAP}~\label{sec:shap}
\begin{figure}[H]
\centering
\includegraphics[width=1\textwidth]{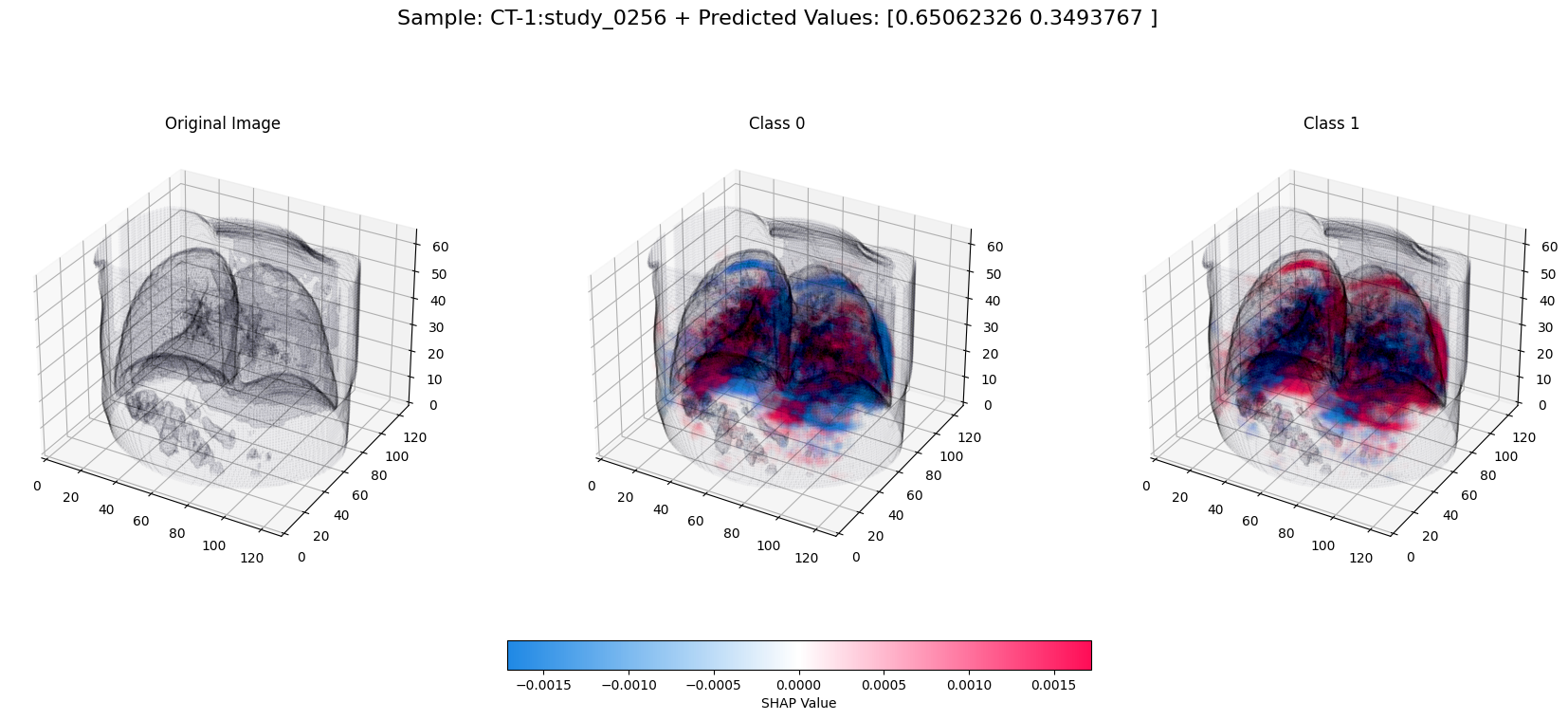}
\caption{SHAP value visualization for sample CT-1. The model predicts Class 0 (no pneumonia) with moderate confidence (0.6506 for Class 0, 0.3494 for Class 1). The left panel shows the original 3D CT scan. The middle panel (Class 0) highlights lung regions with blue areas indicating features pushing the model away from predicting no pneumonia. The right panel (Class 1) shows red regions in the central lungs contributing positively toward the pneumonia prediction, suggesting areas potentially affected by COVID-19.}
\label{fig:shap_ct1}
\end{figure}

As the adoption of AI in medical diagnostics continues to grow, one of the major challenges is the "black-box" nature of neural networks, which makes it difficult to understand how a model reaches its predictions \cite{hamilton2023usingshapvaluesmachine, zeng2024enhancinginterpretabilityshapvalues}. In the context of diagnosing COVID-19 pneumonia from 3D CT scans, the ability to not only classify but also explain the decision-making process behind these classifications is crucial for clinical adoption. To address this need for transparency, we have incorporated SHAP (SHapley Additive exPlanations) values into our model framework, aiming to enhance interpretability. SHAP values provide a principled way to break down the output of a machine learning model by attributing the contribution of each input feature to the final prediction \cite{lundberg2017unifiedapproachinterpretingmodel}. In the case of CT scan classification, SHAP values allow us to quantify the importance of each pixel or voxel within the scan, showing precisely which regions of the image contributed the most to the model’s decision to classify a patient as COVID-19 positive or negative. This is particularly important in healthcare, where clinicians need to understand not only what the model predicts, but also why it predicts it. By implementing SHAP values for 3D, we enable our deterministic neural networks to go beyond just producing a classification output. These models can now offer insight into the specific features driving each prediction, providing a level of explainability that is essential in a medical setting. For instance, if a model classifies a CT scan as COVID-19 positive, SHAP values can highlight the regions in the lungs that were most influential in reaching that conclusion, which may correspond to areas of opacification or other indicative patterns of pneumonia. This added transparency enhances trust in AI-driven diagnostic tools and supports clinicians in validating and interpreting model outputs within the broader clinical context. In addition to enhancing trust, the use of SHAP values in this project also paves the way for better integration of AI models into real-world medical practice. By offering a detailed and interpretable map of the decision-making process, SHAP values empower healthcare professionals to make informed decisions, validate model predictions against their own expertise, and potentially uncover new insights into the imaging patterns associated with COVID-19 pneumonia. This makes SHAP an invaluable tool in bridging the gap between model predictions and clinical understanding, ultimately contributing to more reliable and explainable AI-based healthcare solutions. For our analysis, we selected two sample cases—CT-1 and CT-4—representing different scenarios to demonstrate how the model arrives at its decisions for Class 0 and Class 1.\\
\textbf{first case (CT-1)}. In Fig.~\ref{fig:shap_ct1}, the model predicted Class 0 with moderate confidence, with probabilities of 0.6506 for Class 0 and 0.3494 for Class 1. The SHAP value visualization for Class 0 reveals several blue regions, particularly along the lung periphery, suggesting these areas negatively affect the model's confidence in this prediction. These blue areas indicate the presence of features more consistent with pneumonia, which conflict with the "no pneumonia" classification. In contrast, the Class 1 (pneumonia) panel highlights red regions in the central lung areas, where features such as ground-glass opacities or consolidations could be present. These regions contribute positively to the pneumonia prediction, but they are not strong enough to fully shift the model's decision towards Class 1.\\

\begin{figure}[H]
\centering
\includegraphics[width=1\textwidth]{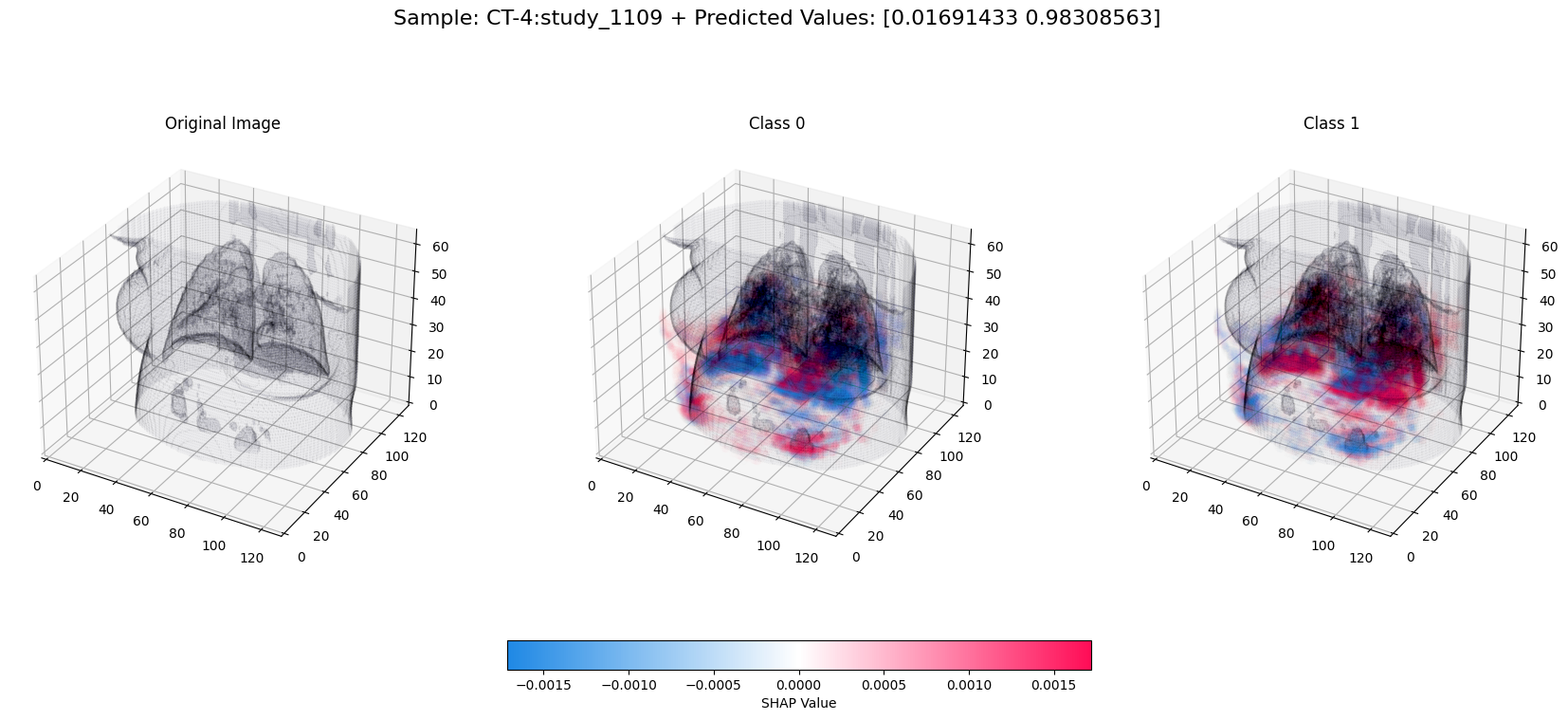}
\caption{SHAP value visualization for sample CT-4. The model is highly confident in predicting Class 1 (pneumonia) with probabilities of 0.0169 for Class 0 and 0.9831 for Class 1. The left panel shows the original 3D CT scan. The middle panel (Class 0) shows widespread blue areas, indicating minimal support for a no pneumonia prediction. The right panel (Class 1) displays dominant red regions, especially in the lower lobes, strongly contributing to the pneumonia classification, consistent with typical manifestations of COVID-19 pneumonia.}
\label{fig:shap_ct4}
\end{figure}

\textbf{Second case (CT-4)}. Fig.~\ref{fig:shap_ct4} shows how the model is highly confident in predicting pneumonia, with a probability of 0.9831 for Class 1 and only 0.0169 for Class 0. The SHAP value visualization for Class 0 shows widespread blue areas throughout the scan, indicating that almost the entire lung is pushing the model away from predicting no pneumonia. These regions show very few features typical of healthy lungs or lungs without pneumonia. In the Class 1 (pneumonia) panel, there are extensive red regions, particularly in the lower lobes, which are strongly contributing to the pneumonia diagnosis. The red shading in these areas suggests the presence of significant abnormal features, likely dense opacities or other infiltrates, that are commonly associated with COVID-19 pneumonia. This widespread presence of abnormal features explains the model's strong confidence in classifying this scan as pneumonia.

Through this SHAP value analysis, we gained a better understanding of how the model interprets specific features in 3D CT scans. These visualizations not only enhance the interpretability of the neural network’s predictions but also shed light on the spatial regions of the lungs that are most influential in driving classification decisions. This level of transparency is crucial for building trust in AI models for medical diagnosis, as it allows clinicians to cross-reference model predictions with observable imaging features, thereby supporting more informed and reliable clinical decision-making.
\section{Discussion}~\label{sec8:conc}
One notable advantage of Bayesian neural networks is their ability to provide a probabilistic interpretation of predictions, allowing for the quantification of uncertainty. This feature is crucial for clinical decision-making, as it empowers clinicians to make more informed decisions, particularly in cases where the consequences of false positives or false negatives are significant. However, there are also several disadvantages to consider. Bayesian methods typically require more computational resources and time compared to deterministic counterparts due to the need for sampling-based inference techniques.  Moreover, the effectiveness of Bayesian neural networks heavily relies on the availability of large and diverse datasets for training, which may be limited, especially in medical imaging. This clearly create an scenario where the training can be challenging and time-consuming.   Besides, Deep learning models, especially complex neural networks proposed in this work, can be difficult to interpret, making it challenging to understand the reasons behind their decisions. This can be particularly problematic in medical applications where transparency and accountability are crucial. Shap values allow to reduce the opacity of interpretability of the models, but in scenarios where the amount of classes increases, this technique provide complex responses that make the analysis even harder.

\section{Summary and conclusions}
Accurately classifying COVID-19 pneumonia in 3D CT scans remains a significant challenge in medical image analysis. While deterministic neural networks have shown promise, their point estimate outputs limit their clinical utility. These models often lack the ability to quantify uncertainty, which is crucial for reliable diagnostic decision-making. This study has demonstrated the efficacy of deep learning techniques in addressing complex medical imaging challenges. By carefully selecting optimal pixel alternatives and fine-tuning hyperparameters, we have established a robust foundation for training accurate and reliable classification models. The integration of Bayesian neural networks has proven pivotal in quantifying uncertainty within model predictions. This capability is essential for reliable and interpretable clinical decision-making, especially in high-stakes medical contexts. Furthermore, the calibration analysis has provided valuable insights into the confidence and reliability of our models. By understanding the alignment between predicted probabilities and actual outcomes, we can enhance the trustworthiness of our predictions. The exploration of uncertainty through confidence intervals has highlighted the importance of assessing the reliability of model predictions. This knowledge is critical for making informed clinical decisions and mitigating potential risks associated with inaccurate predictions. Furthermore, explainability has been also a powerful tool built  in this work to interpret the results and understand which regions the model focuses on to make decision. Overall, this study underscores the potential of deep learning to revolutionize medical imaging and improve patient care. By applying these advanced techniques, we can unlock new insights, enhance diagnostic accuracy, and ultimately contribute to better health outcomes.
\appendix

\section{Architecture}
\begin{figure}[H]
    \centering
    \includegraphics[width=1\textwidth]{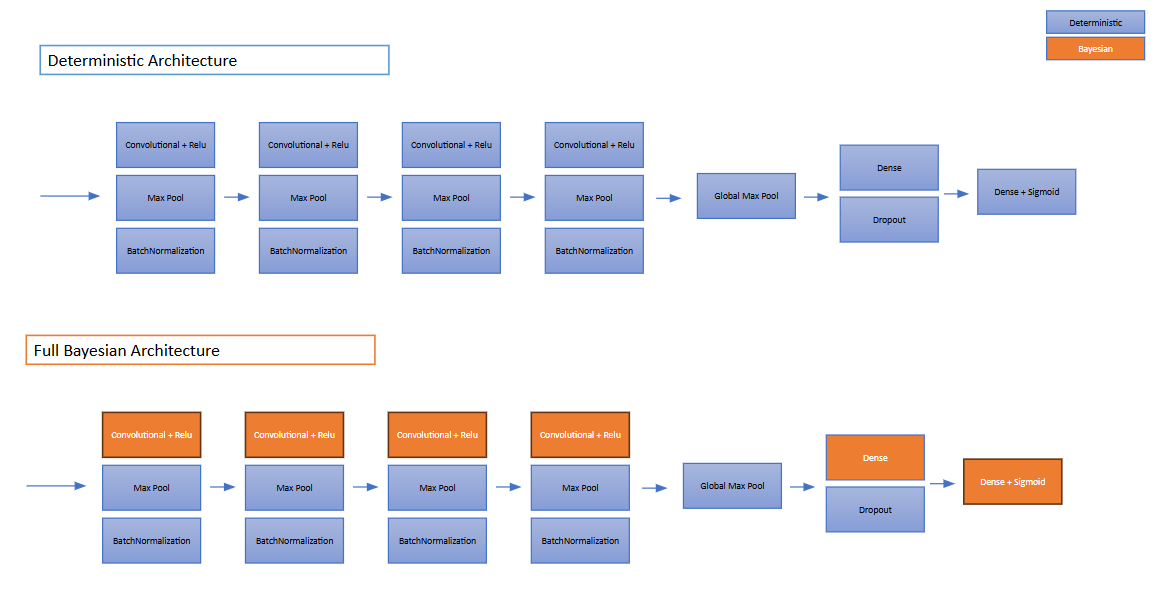}
    \caption{Final architectures of the models}
    \label{fig:architecturedet}
\end{figure}

Fig.~\ref{fig:architecturedet} depicts the architecture used in the manuscript for the deterministic and its Bayesian counterpart.  

\section{Metrics of all Deterministic Models} \label{anexo_tabla_modelos_1}
In table~\ref{tab:deterministas_test_todos} reports the metrics for the entire set of deterministic models selected in the paper.
\begin{table}[H]
\tiny  
\centering
\begin{tabular}{|l|l|l|l|l|l|l|}
\hline
\textbf{model\_name} & \textbf{accuracy} & \textbf{precision} & \textbf{recall} & \textbf{f1} & \textbf{Cohens kappa} & \textbf{ROC AUC} \\
\hline
DNN\_W1\_V1 & 0.845238 & 0.956522 & 0.647059 & 0.771930 & 0.661290 & 0.920588 \\
DNN\_W1\_V2 & 0.642857 & 0.642857 & 0.264706 & 0.375000 & 0.181818 & 0.740000 \\
DNN\_W2\_V1 & 0.892857 & 0.878788 & 0.852941 & 0.865672 & 0.776596 & 0.956471 \\
DNN\_W3\_V1 & 0.857143 & 0.805556 & 0.852941 & 0.828571 & 0.706294 & 0.956471 \\
DNN\_W4\_V1 & 0.904762 & 0.825000 & 0.970588 & 0.891892 & 0.807780 & 0.974118 \\
Resnet18\_W4\_V4 & 0.750000 & 0.644444 & 0.852941 & 0.734177 & 0.506711 & 0.837059 \\
Resnet34\_W4\_V5 & 0.785714 & 0.766667 & 0.676471 & 0.718750 & 0.546763 & 0.865882 \\
Seresnet18\_W4\_V6 & 0.761905 & 0.850000 & 0.500000 & 0.629630 & 0.471033 & 0.842941 \\
Seresnet34\_W4\_V7 & 0.773810 & 0.894737 & 0.500000 & 0.641509 & 0.494937 & 0.871176 \\
Resnet18\_W4\_V8 & 0.642857 & 0.625000 & 0.294118 & 0.400000 & 0.190231 & 0.774118 \\
Resnet18\_W4\_V9 & 0.726190 & 0.789474 & 0.441176 & 0.566038 & 0.388608 & 0.742353 \\
Resnet18\_W4\_V10 & 0.714286 & 0.678571 & 0.558824 & 0.612903 & 0.389831 & 0.803529 \\
Resnet18\_W4\_V11 & 0.761905 & 0.750000 & 0.617647 & 0.677419 & 0.491525 & 0.834118 \\
Resnet18\_W4\_V12 & 0.690476 & 0.750000 & 0.352941 & 0.480000 & 0.298201 & 0.778824 \\
Resnet18\_W4\_V13 & 0.619048 & 0.520000 & 0.764706 & 0.619048 & 0.264770 & 0.740588 \\
Resnet34\_W4\_V14 & 0.678571 & 0.652174 & 0.441176 & 0.526316 & 0.296526 & 0.734118 \\
Seresnet18\_W4\_V15 & 0.845238 & 0.838710 & 0.764706 & 0.800000 & 0.674224 & 0.840588 \\
Seresnet34\_W4\_V16 & 0.726190 & 0.923077 & 0.352941 & 0.510638 & 0.369452 & 0.812941 \\
DNN\_W4\_V1\_Confirmation & 0.892857 & 0.857143 & 0.882353 & 0.869565 & 0.778689 & 0.962941 \\
Efficientnetb0\_W4\_V17 & 0.607143 & 0.518519 & 0.411765 & 0.459016 & 0.156934 & 0.616471 \\
Densenet121\_W4\_V18 & 0.738095 & 0.657895 & 0.735294 & 0.694444 & 0.466513 & 0.811765 \\
Resnet18\_W1\_V4 & 0.773810 & 0.800000 & 0.588235 & 0.677966 & 0.509828 & 0.872353 \\
Resnet18\_W1\_V5 & 0.797619 & 0.869565 & 0.588235 & 0.701754 & 0.557072 & 0.894706 \\
Resnet18\_W1\_V6 & 0.738095 & 0.676471 & 0.676471 & 0.676471 & 0.456471 & 0.822353 \\
DNN\_W4\_V1\_Optimized & 0.964286 & 0.918919 & 1.000000 & 0.957746 & 0.926914 & 0.989412 \\
DNN\_W4\_V1\_Hyper\_Reg & 0.892857 & 0.804878 & 0.970588 & 0.880000 & 0.784738 & 0.981176 \\
\hline
\end{tabular}
\caption{Metrics of all Deterministic Models.}
\label{tab:deterministas_test_todos}
\end{table}

\section{Description of evaluated models} \label{anexo_tabla_modelos}

The tables related in this annex briefly describe the details of each model implemented during the project.

{\tiny
\begin{longtable}
{|p{0.2\linewidth}|p{0.08\linewidth}|p{0.08\linewidth}|p{0.15\linewidth}|p{0.05\linewidth}|p{0.07\linewidth}|p{0.07\linewidth}|p{0.07\linewidth}|}

\caption{3D Models Description} \label{tab:info_modelos_3D} \\
\hline
\textbf{Model Name} & \textbf{Dimension} & \textbf{Type} & \textbf{Description} & \textbf{Window} & \textbf{Learning Rate} & \textbf{Epochs} & \textbf{Batch Size} \\ \hline
\endfirsthead

\multicolumn{8}{c}%
{{\tablename\ \thetable{} -- Continued from previous page}} \\
\hline
\textbf{Model Name} & \textbf{Dimension} & \textbf{Type} & \textbf{Description} & \textbf{Window} & \textbf{Learning Rate} & \textbf{Epochs} & \textbf{Batch Size} \\ \hline
\endhead

\hline \multicolumn{8}{r}{{Continued on the next page}} \\
\endfoot

\hline
\endlastfoot

DNN\_W1\_V1 & 3D & Deterministic & Model with architecture taken from the Keras exercise for 3D neural networks. & W1 & 0,001 & 100 & 2 \\ \hline
DNN\_W1\_V2 & 3D & Deterministic & Model with architecture taken from the Keras exercise for 3D neural networks + Implementation of Data Augmentation techniques & W1 & 0,0001 & 100 & 2 \\ \hline
Resnet18\_W1\_V4 & 3D & Deterministic & Implementing the Resnet18 model from the classification\_models\_3D package and adding a GlobalMaxPooling3D layer & W1 & 0,0001 & 100 & 2 \\ \hline
Resnet18\_W1\_V5 & 3D & Deterministic & Implementing the Resnet18 model from the classification\_models\_3D package and adding a GlobalMaxPooling3D layer & W1 & 0,001 & 100 & 2 \\ \hline
Resnet18\_W1\_V6 & 3D & Deterministic & Implementing the Resnet18 model from the classification\_models\_3D package and adding a GlobalAveragePooling3D layer & W1 & 0,0001 & 100 & 2 \\ \hline
DNN\_W2\_V1 & 3D & Deterministic & Model with architecture taken from the Keras exercise for 3D neural networks. & W2 & 0,0001 & 100 & 2 \\ \hline
DNN\_W3\_V1 & 3D & Deterministic & Model with architecture taken from the Keras exercise for 3D neural networks. & W3 & 0,0001 & 100 & 2 \\ \hline
DNN\_W4\_V1 & 3D & Deterministic & Model with architecture taken from the Keras exercise for 3D neural networks. & W4 & 0,0001 & 100 & 2 \\ \hline
DNN\_W4\_V1\_Confirmation & 3D & Deterministic & Model with architecture taken from the Keras exercise for 3D neural networks. (Executed for the 2nd time to confirm values) & W4 & 0,0001 & 100 & 2 \\ \hline
DNN\_W4\_V1\_Optimized & 3D & Deterministic & Model with architecture taken from the Keras exercise for 3D neural networks. + Optimization with Keras-Tuner & W4 & 0,001 & 60 & 2 \\ \hline
DNN\_W4\_V1\_Hyper\_Reg & 3D & Deterministic & Model with architecture taken from the Keras exercise for 3D neural networks. + Optimization with Keras-Tuner in terms of Regularizers & W4 & 0,001 & 60 & 2 \\ \hline
Resnet18\_W4\_V4 & 3D & Deterministic & Resnet18 & W4 & 0,0001 & 100 & 2 \\ \hline
Resnet34\_W4\_V5 & 3D & Deterministic & Resnet34 & W4 & 0,0001 & 100 & 2 \\ \hline
Seresnet18\_W4\_V6 & 3D & Deterministic & Seresnet18 & W4 & 0,0001 & 100 & 2 \\ \hline
Seresnet34\_W4\_V7 & 3D & Deterministic & Seresnet34 & W4 & 0,0001 & 100 & 2 \\ \hline
Resnet18\_W4\_V8 & 3D & Deterministic & Resnet18 + Penultimate layer of the 3D architecture will be GlobalMaxPooling3D & W4 & 0,0001 & 100 & 2 \\ \hline
Resnet18\_W4\_V9 & 3D & Deterministic & Resnet18 + Penultimate layer of the 3D architecture will be GlobalMaxPooling3D + Filters 3D Model: 8 & W4 & 0,0001 & 100 & 2 \\ \hline
Resnet18\_W4\_V10 & 3D & Deterministic & Resnet18 + Penultimate layer of the 3D architecture will be GlobalMaxPooling3D + Filters 3D Model: 16 & W4 & 0,0001 & 100 & 2 \\ \hline
Resnet18\_W4\_V11 & 3D & Deterministic & Resnet18 + Penultimate layer of the 3D architecture will be GlobalMaxPooling3D + Filters 3D Model: 32 & W4 & 0,0001 & 100 & 2 \\ \hline
Resnet18\_W4\_V12 & 3D & Deterministic & Resnet18 + Penultimate layer of the 3D architecture will be GlobalMaxPooling3D + Filters 3D Model: 8 & W4 & 0,0001 & 100 & 6 \\ \hline
Resnet18\_W4\_V13 & 3D & Deterministic & Resnet18 + Penultimate layer of the 3D architecture will be GlobalMaxPooling3D & W4 & 0,0001 & 100 & 4 \\ \hline
Resnet34\_W4\_V14 & 3D & Deterministic & Resnet34 + Penultimate layer of the 3D architecture will be GlobalMaxPooling3D & W4 & 0,0001 & 100 & 4 \\ \hline
Seresnet18\_W4\_V15 & 3D & Deterministic & Seresnet18 + Penultimate layer of the 3D architecture will be GlobalMaxPooling3D & W4 & 0,0001 & 100 & 4 \\ \hline
Seresnet34\_W4\_V16 & 3D & Deterministic & Seresnet34 + Penultimate layer of the 3D architecture will be GlobalMaxPooling3D & W4 & 0,0001 & 100 & 2 \\ \hline
Efficientnetb0\_W4\_V17 & 3D & Deterministic & Efficientnetb0 + Penultimate layer of the 3D architecture will be GlobalMaxPooling3D & W4 & 0,0001 & 100 & 2 \\ \hline
Densenet121\_W4\_V18 & 3D & Deterministic & Densenet121 + Penultimate layer of the 3D architecture will be GlobalMaxPooling3D & W4 & 0,0001 & 100 & 2 \\ \hline
DNN\_W4\_V1\_Uncertainty\_V1 & 3D & Bayesian & Keras (Optimized Tuner) + Bernoulli Layer || IT HAS NO METRICS & W4 & 0,001 & 150 & 2 \\ \hline
DNN\_W4\_V1\_Uncertainty\_V2 & 3D & Bayesian & Keras (Optimized Tuner) + Bernoulli Layer (MEAN) || IT HAS NO METRICS & W4 & 0,001 & 150 & 2 \\ \hline
DNN\_W4\_V1\_Uncertainty\_V2\_2 & 3D & Bayesian & Keras (Optimized Tuner) + Bernoulli Layer (MEAN) & W4 & 0,0001 & 300 & 2 \\ \hline
DNN\_W4\_V1\_Uncertainty\_V3 & 3D & Bayesian & Keras (Optimized Tuner) + MNFDense Bernoulli Layer || IT HAS NO METRICS & W4 & 0,001 & 150 & 2 \\ \hline
DNN\_W4\_V1\_Uncertainty\_V4 & 3D & Bayesian & Keras (Optimized Tuner) + MNFDense Bernoulli Layer (MEAN) || IT HAS NO METRICS & W4 & 0,001 & 150 & 2 \\ \hline
DNN\_W4\_V1\_Uncertainty\_V4\_2 & 3D & Bayesian & Keras (Optimized Tuner) + MNFDense Bernoulli Layer (MEAN) & W4 & 0,0001 & 800 & 2 \\ \hline
DNN\_W4\_V1\_Uncertainty\_V4\_2C & 3D & Bayesian & Keras (Optimized Tuner) + MNFDense Bernoulli Layer (MEAN) | It's the same V4\_2 model, just run again because the saved weights from the original were corrupted & W4 & 0,0001 & 800 & 2 \\ \hline
DNN\_W4\_V1\_Uncertainty\_V5 & 3D & Bayesian & Keras (Optimized Tuner) + DenseFlipout Bernoulli Layer (MEAN) & W4 & 0,0001 & 300 & 2 \\ \hline
DNN\_W4\_V1\_Uncertainty\_V6 & 3D & Bayesian & Keras (Optimized Tuner) + DenseLocalReparameterization Bernoulli Layer (MEAN) & W4 & 0,0001 & 300 & 2 \\ \hline
DNN\_W4\_V1\_Uncertainty\_V7 & 3D & Bayesian & Keras (Optimized Tuner) + Bernoulli DenseReparameterization Layer (MEAN) & W4 & 0,0001 & 300 & 2 \\ \hline
DNN\_W4\_V1\_Uncertainty\_V4\_3 & 3D & Bayesian & Keras (Optimized Tuner) + MNFDense Bernoulli Layer (MEAN) + MNFConv3D Layers & W4 & 0,0001 & 300 & 2 \\ \hline
DNN\_W4\_V1\_Uncertainty\_V4\_3C & 3D & Bayesian & Keras (Optimized Tuner) + MNFDense Bernoulli Layer (MEAN) + MNFConv3D Layers | It's the same V4\_3 model, just run again because the saved weights from the original were corrupted & W4 & 0,0001 & 300 & 2 \\ \hline
DNN\_W4\_V1\_Uncertainty\_MCDropout & 3D & Bayesian & Keras (Optimized Tuner) + Dropout MC & W4 & 0,0001 & 100 & 2 \\ \hline
DNN\_W4\_V1\_Uncertainty\_V8 & 3D & Bayesian & Keras (Optimized Tuner) + DenseFlipout Bernoulli Layer (MEAN) + ConvFlipout Layers & W4 & 0,0001 & 300 & 2 \\ \hline
DNN\_W4\_V1\_Uncertainty\_V9 & 3D & Bayesian & Keras (Optimized Tuner) + Bernoulli DenseReparameterization Layer (MEAN) + Conv Reparameterization Layers & W4 & 0,0001 & 300 & 2 \\ \hline

\end{longtable}
}

\bibliographystyle{elsarticle-harv} 
\bibliography{example}






\end{document}